\documentclass[twocolumn]{aastex631}
\usepackage{graphicx}
\usepackage{float}
\usepackage{amsmath}
\usepackage{upgreek}

\newcommand{\julia}{\texttt{julia}}
\newcommand{\angstrom}{$\rm \mathring A$}

\revised{\today}
\shortauthors{Su et al.}

\begin{document}

\title{A new timescale-mass scaling for the optical variation of active galactic nuclei across the intermediate-mass to supermassive scales}

\correspondingauthor{Zhen-Yi Cai}
\email{zcai@ustc.edu.cn}

\author[0000-0001-8515-7338]{Zhen-Bo Su}
\affiliation{CAS Key Laboratory for Research in Galaxies and Cosmology, Department of Astronomy, University of Science and Technology of China, Hefei 230026, China}
\affiliation{School of Astronomy and Space Science, University of Science and Technology of China, Hefei 230026, China}

\author[0000-0002-4223-2198]{Zhen-Yi Cai}
\affiliation{CAS Key Laboratory for Research in Galaxies and Cosmology, Department of Astronomy, University of Science and Technology of China, Hefei 230026, China}
\affiliation{School of Astronomy and Space Science, University of Science and Technology of China, Hefei 230026, China}

\author[0000-0002-0771-2153]{Mouyuan Sun}
\affiliation{Department of Astronomy, Xiamen University, Xiamen 361005, China}

\author[0000-0001-8416-7059]{Hengxiao Guo}
\affiliation{Shanghai Astronomical Observatory, Chinese Academy of Sciences, 80 Nandan Road, Shanghai 200030, China}

\author[0000-0003-3137-1851]{Wei-Min Gu}
\affiliation{Department of Astronomy, Xiamen University, Xiamen 361005, China}

\author[0000-0002-4419-6434]{Jun-Xian Wang}
\affiliation{CAS Key Laboratory for Research in Galaxies and Cosmology, Department of Astronomy, University of Science and Technology of China, Hefei 230026, China}
\affiliation{School of Astronomy and Space Science, University of Science and Technology of China, Hefei 230026, China}


\begin{abstract}

Variability of active galactic nuclei (AGNs) has long been servicing as an essential avenue of exploring the accretion physics of black hole (BH). There are two commonly used methods for analyzing AGN variability. First, the AGN variability, characterized by the structure function (SF) of a single band, can be well described by a damped random walk (DRW) process on timescales longer than $\sim$weeks, shorter than which departures have been reported. Second, the color variation (CV) between two bands behaves timescale-dependent, raising challenges to the widely accepted reprocessing scenario. However, both the departure from the DRW process and the timescale-dependent CV are hitherto limited to AGNs, mainly quasars, at the supermassive scale. Here, utilizing the high-cadence multi-wavelength monitoring on NGC 4395 harboring an intermediate-mass BH, we unveil at the intermediate-mass scale for the first time, prominent departures from the DRW process at timescales shorter than $\sim$hours in all three nights and bands, and plausible timescale-dependent CVs in the two longest nights of observation. Furthermore, comparing SFs of NGC 4395 to four AGNs at the supermassive scale, we suggest a new scaling relation between the timescale ($\uptau$; across nearly three orders of magnitude) and the BH mass ($M_{\rm BH}$): $\uptau \propto M^\gamma_{\rm BH}$ where the exponent $\gamma$ is likely $\simeq 0.6 - 0.8$. This exponent differs from most previous measurements, but confirms a few and is consistent with a recent theoretical prediction, suggesting a similar accretion process in AGNs across different mass scales. 

\end{abstract}



\section{Introduction} \label{sec:intro}

The ultraviolet-to-optical (UV/optical) continuum emission of active galactic nuclei (AGNs), which encompass luminous quasars and their low-luminosity counterparts, is believed to originate from the accretion disk surrounding the central massive black hole (BH). Since the angular sizes of the majority of AGNs are too small to be spatially resolved by the current best telescopes, studying AGN variability in UV/optical provides an alternative avenue of unveiling the physics of the AGN central engine. 

Hitherto, many properties of AGN variability have been found. In the UV/optical, the stochastic AGN variability can be well characterized by a damped random walk (DRW) process as least on timescales of weeks to years \citep[e.g., ][]{Kelly2009ApJ...698..895K,Zu2013ApJ...765..106Z,Kozlowski2016ApJ...826..118K}. Then correlations between the DRW parameters (i.e., the de-correlation or damping timescale, $\tau_{\rm d}$, and the asymptotic variability amplitude on long timescales, ${\rm SF}_\infty$) and the physical properties of AGNs, such as the BH mass, the Eddington ratio, and the luminosity, have been explored 
\citep[e.g., ][]{Kelly2009ApJ...698..895K,Kozlowshi2010ApJ...708..927K,MacLeod2010ApJ...721.1014M,caplar2017, Sun2018ApJ...866...74S,Suberlak2021ApJ...907...96S,Burke2021Sci...373..789B,tang2023NatAs...7..473T, Arevalo2023}. 
However, departures from the DRW process have been reported at timescales shorter than $\sim$weeks \citep[e.g.,][]{Mushotzky2011ApJ...743L..12M,Kelly2014,Kasliwal2015,Simm2016A&A...585A.129S}, probably at timescales longer than years \citep{Guo2017ApJ...847..132G}, and at shorter extreme UV wavelengths \citep{Zhu2016ApJ...832...75Z}. Note that these known departures from the DRW process are only found in quasars and AGNs with typically BHs massive than $\sim 10^7~M_\odot$.

In addition, by analyzing variations between UV/optical continuum bands, the variation at shorter wavelength (bluer band) is usually larger than that at longer wavelength (redder band), which is widely known as the bluer-when-brighter (BWB) color variation (CV) since \citeauthor{Cutri1985ApJ...296..423C} (\citeyear{Cutri1985ApJ...296..423C}; see also, e.g., \citealt{Schmidt2012ApJ...744..147S,Zuo2012ApJ...758..104Z,Ruan2014ApJ...783..105R,Guo2016ApJ...822...26G}). Understanding on the CV had not been refurbished until \citet{sun2014ApJ...792...54S} who discovered that the ensemble CV of the Sloan Digital Sky Survey (SDSS) Stripe 82 quasars is timescale-dependent, that is, the CV is more prominent at shorter timescales. The timescale-dependent CV immediately demonstrates that the origin of the CV is neither attributed to a simple combination of a variable AGN with a fixed bluer spectral slope and an invariable redder host galaxy \citep{sun2014ApJ...792...54S}, nor to changes in the global accretion rates \citep{Cai2016}. Instead, it suggests that the CV is intrinsic to the accretion disk, and furthermore supports an origin of the thermal fluctuation for the UV/optical variation in quasars (\citealt{Cai2016}; see also \citealt{Kelly2009ApJ...698..895K,Dexter2011ApJ...727L..24D}).

The power of analyzing the timescale-dependent CV is striking. By extending the color analysis on a quasar sample observed at shorter rest-frame wavelengths from near UV to extreme UV, \citet{Zhu2016ApJ...832...75Z} not only confirmed the timescale dependence in the UV but also unveiled that the thermal fluctuation in the inner disk may deviate from the DRW process. Moreover, comparing the CV difference between the radio-loud and radio-quiet quasars, \cite{Cai2019} suggested that the inner disks in the radio-loud quasars may be more stable, providing new clues to the jet launching. From quasar samples to individual AGNs, the role of analyzing the timescale-dependent CV is also attractive. \cite{Zhu2018} successfully identified for the first time the UV/optical timescale-dependent CV in an individual AGN, i.e., NGC 5548 (at $z = 0.017175$ and with $M_{\rm BH} \sim 5 \times 10^7~M_\odot$; \citealt{Pancoast2014MNRAS.445.3073P}), challenging the traditional thermal reprocessing scenario, where the UV/optical variation is assumed to be the thermal reprocessing of the variable X-ray emission. In another individual AGN, i.e., NGC 4051 (with $M_{\rm BH} \sim 2 \times 10^6~M_\odot$; \citealt{Denney2009ApJ...702.1353D}), \citet{Wu2020SCPMA..6329512W} found that the timescale-dependent CV between two hard X-ray bands hints at random individual solar-like flares perturbing the global variation of the extended corona. 

Unfortunately, the timescale-dependent CV has not gotten enough attention by the community even we have progressively developed the method and have applied it to more and more situations aforementioned over the past decade. Furthermore, the timescale-dependent CV, as well as the aforementioned departure from the DRW process, are hitherto found in quasars or AGNs whose BH masses are larger than $10^{6-7}~M_\sun$.

Recently, exploring the optical variation properties of AGNs \citep{Burke2020ApJ...899..136B,Montano2022, McHardy2023MNRAS.519.3366M} has been extended to a nearby dwarf galaxy, NGC 4395 \citep[at a redshift of $z = 0.001064$;][]{Filippenko1989,denBrok2015ApJ...809..101D}, harboring an intermediate-massive BH (IMBH) with a BH mass of $\sim 10^{4-5}~M_\odot$ \citep[e.g.,][]{Filippenko&Ho2003ApJ...588L..13F, Peterson2005ApJ...632..799P, Woo2019NatAs...3..755W, Cho2021ApJ...921...98C}. On one hand, using $\sim$month-long and 30 minute-cadence observations of the Transiting Exoplanet Survey Satellite (TESS), \cite{Burke2020ApJ...899..136B} found that the optical variation of NGC 4395 can also be described by the same DRW process as the luminous quasars. On the other hand, both \citet{Montano2022} and \citet{McHardy2023MNRAS.519.3366M} mainly focus on the inter-band lags. 

Here, benefiting from these recent very high cadence (as short as several seconds to several minutes) and simultaneous multi-band monitoring campaigns (\citealt{Montano2022, McHardy2023MNRAS.519.3366M}), we revisit the optical variation properties of NGC 4395 and unveil at the intermediate-mass scale for the first time a prominent departure from the DRW process and a plausible timescale-dependent CV.
Section~\ref{sec:data} successively introduces the multi-band light curves of NGC 4395, the structure functions (SFs), and the CVs, complemented with the methodology for automatically determining a proper range of time intervals where results on the SF or CV are not significantly affected by either the photometric uncertainty at short timescales or the limited length of the light curve at long timescales. Discussions are presented in Section~\ref{sec:dis}, followed by a brief summary in Section~\ref{sec:conclusion}.

The codes used in this work are available on Zenodo: \dataset[doi:10.5281/zenodo.10428783]{https://doi.org/10.5281/zenodo.10428783}.

\section{Optical Variation Properties of NGC 4395}\label{sec:data}

\subsection{Light Curves}\label{sec:lc}

\begin{figure*}
    \centering
    \includegraphics[width=0.8\textwidth]{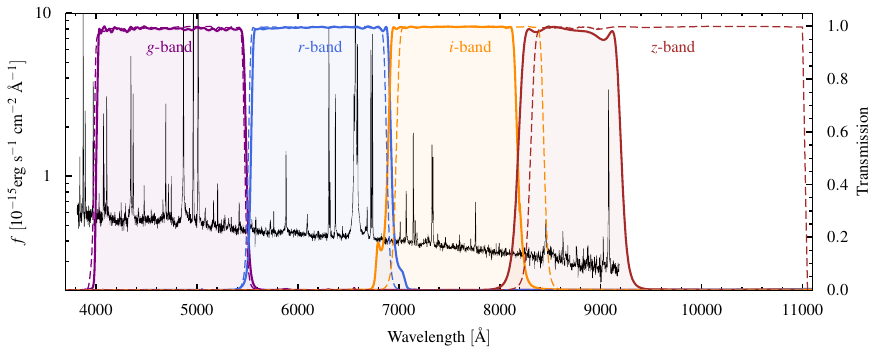}
    \caption{The FNT (colored solid curves with shaded regions) and GTC (colored dashed curves) $griz$-band transmission curves (right $y$-axis) on top of an SDSS spectrum for NGC 4395 (black; left $y$-axis).}
    \label{fig:trans_spec}
\end{figure*}

Using a four-channel MuSCAT3 camera on the 2 m Faulkes North Telescope (FNT) simultaneously observing in $g'$, $r'$, $i'$, and $z'_s$ bands\footnote{The central wavelengths of the $g'$, $r'$, $i'$, and $z'_s$ bands are 4770~\angstrom, 6215~\angstrom, 7545~\angstrom, and 8700~\angstrom, respectively \citep[][cf. their Table~2]{Montano2022}. See \url{https://lco.global/observatory/instruments/filters} for filters and Figure~\ref{fig:trans_spec} for an illustration.}, \citet{Montano2022} monitored NGC 4395 in two consecutive nights: $\sim6.7$ hr on 2022 April 26 (hereafter, the FNT-1) and $\sim6.2$ hr on 2022 April 27 (hereafter, the FNT-2). The typical cadences in the $g'$, $i'$, and $z'_s$ bands are 103.68~s, while denser cadences of 34.56~s are found in the $r'$ band. Since analyzing the CV requires quasi-simultaneous observations and instants among these four bands are not exactly the same, we re-bin all four light curves in the same bin sizes of 103.68~s. Within a bin, multiple flux points, found primarily in the $r'$ band, are averaged and the associated errors are propagated accordingly.

Similarly, NGC 4395 was simultaneously observed in $u_s$, $g_s$, $r_s$, $i_s$, and $z_s$ bands\footnote{The effective wavelengths of the $g_s$, $r_s$, $i_s$, and $z_s$ bands are 4778~\angstrom, 6201~\angstrom, 7640~\angstrom, and 9066~\angstrom, respectively \citep[][cf. their Table~3]{McHardy2023MNRAS.519.3366M}. See \url{http://www.vikdhillon.staff.shef.ac.uk/ultracam/filters/filters.html} for filters and Figure~\ref{fig:trans_spec} for an illustration.} by \citet{McHardy2023MNRAS.519.3366M} for 6 hr from 2018 April 16 to 17 with HiPERCAM on the 10.4 m Gran Telescopio Canarias (GTC). The cadences are as short as 3~s for the $g_s$, $r_s$, $i_s$, and $z_s$ bands, and 15~s for the $u_s$ band. In order to compare the variation properties between \citet{McHardy2023MNRAS.519.3366M} and \citet{Montano2022}, we exclude the $u_s$ band. Moreover, we only consider the light curves within the first 8000~s as strongly suggested by \citet[][affected by the loss of tracking as a result of the motion of NGC 4395 and the calibration stars over the CCDs]{McHardy2023MNRAS.519.3366M}. 
The AGN magnitudes have been derived combining the count-rate ratios of the AGN to Star~1 with the SDSS magnitudes of star~1 (cf. Table~2 of \citealt{McHardy2023MNRAS.519.3366M}).
Note that the flux calibration may not be particularly accurate since Star~1 might have varied slightly since measured by SDSS (I. McHardy 2023, private communication).

\begin{figure*}[t!]
    \centering
    \includegraphics[width=0.8\textwidth]{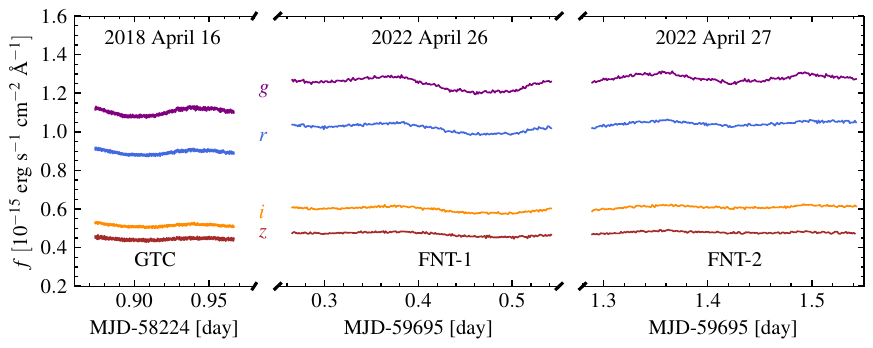}
    \caption{Recent very high cadence and multi-band ($g$, $r$, $i$, and $z$ from top to bottom, respectively) light curves of NGC 4395 observed in three nights on 2018 April 16 \citep[2.2 hr;][]{McHardy2023MNRAS.519.3366M} and on 2022 April 26 to 27 \citep[6.7 hr and 6.2 hr;][]{Montano2022}. The $g$-, $r$-, $i$-, and $z$-band mean fluxes of \citet{McHardy2023MNRAS.519.3366M} are slightly smaller than those of \citet[][average over two nights]{Montano2022} by factors of $\simeq 1.15$, $\simeq 1.15$, $\simeq 1.17$, and $\simeq 1.07$, respectively (Table~\ref{tab:fracs}). The flux densities are in units of $\rm 10^{-15}~erg~s^{-1}~cm^{-2}~\text{\AA}^{-1}$.}
    \label{fig:lc}
\end{figure*}

\begin{figure*}[t!]
    \centering
    \includegraphics[width=0.9\textwidth]{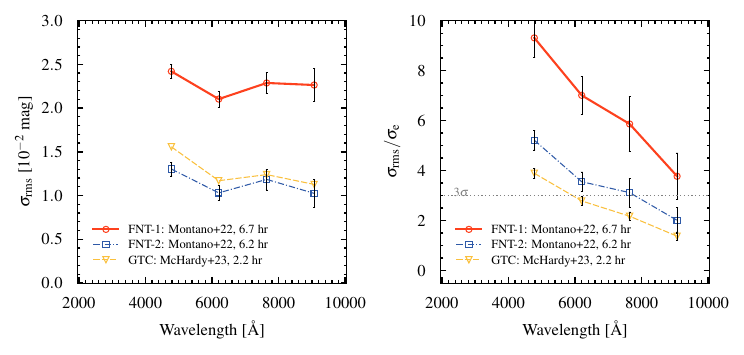}
    \caption{Left panel: the excess variance, $\sigma_{\rm rms}$, as a function of wavelength for NGC 4395 monitored in three nights (Table~\ref{tab:fracs}): FNT-1 (open circles), FNT-2 (open squares), and GTC (open triangles).
    Right panel: the significance of variation, $\sigma_{\rm rms}/\sigma_{\rm e}$, as a function of wavelength, where $\sigma_e$ is the mean photometric uncertainties in magnitude (Table~\ref{tab:fracs}). The dotted horizontal line indicates a 3$\sigma$ level.}
    \label{fig:fvar}
\end{figure*}

Figure \ref{fig:trans_spec} displays the FNT and GTC transmission curves on top of an SDSS spectrum ($\rm MJD = 53819$, $\rm PlateID = 2015$, and $\rm FiberID = 251$) for NGC 4395. Since the filter differences are small (except the $z'_s$/$z_s$ band), we simply use the same abbreviation $griz$ hereafter.
Figure \ref{fig:lc} illustrates the $griz$-band light curves of NGC 4395 in three nights used in this work, whose flux densities are in units of $\rm erg~s^{-1}~cm^{-2}~\text{\AA}^{-1}$. Note that differences in the mean flux densities between \citet{Montano2022} and \citet{McHardy2023MNRAS.519.3366M} are not large, i.e., $\lesssim 17\%$. Table~\ref{tab:fracs} tabulates the mean flux density, the mean photometric uncertainty ($\sigma_{\rm e}$), and the variation amplitude for each band and night.
Following \citet{Vaughan2003MNRAS}, the variation amplitude in each band is characterized by the excess variation, $\sigma_{\rm rms}$. Figure~\ref{fig:fvar} illustrates $\sigma_{\rm rms}$ as well as the significance of variation, $\sigma_{\rm rms}/\sigma_{\rm e}$, as a function of wavelength.

\begin{deluxetable*}{ccccccccc}
\tablewidth{0pt}
\tablecaption{Variation properties of NGC 4395 and the best-fit parameters for its SFs.}\label{tab:fracs}
\tablehead{\colhead{Date} & \colhead{Band} & \colhead{$\bar f^x$} & $\sigma_{\rm rms}$ & \colhead{$\sigma^x_{\rm e}$}  & \colhead{$\sigma_n$} & \colhead{$\beta$} & \colhead{$\tau_{\rm d}$} & \colhead{$\rm SF_{\infty}$} \\
& & & ($10^{-2}$ mag) & \colhead{($10^{-2}$ mag)} & \colhead{($10^{-2}$ mag)} & & \colhead{($10^3$ s)}&\colhead{($10^{-2}$ mag)} \\
\colhead{(1)} & \colhead{(2)} & \colhead{(3)} & \colhead{(4)} & \colhead{(5)} & \colhead{(6)} & \colhead{(7)} & \colhead{(8)} & \colhead{(9)}
}
\startdata
2018 April 16 & \emph{g} & 1.10 & $1.56 \pm 0.07$ & $0.40\pm0.01$ & $0.38\pm0.01$  &  $1.79\pm0.02$ & $2.5\pm0.1$ & $3.7\pm0.1$ \\
GTC & \emph{r} & 0.89 &  $1.17 \pm 0.07$ & $0.42\pm0.01$  &  $0.39\pm0.01$  &  $1.84\pm0.03$ & $2.2\pm0.1$  &  $2.8\pm0.1$ \\
 & \emph{i} & 0.52 & $1.24 \pm 0.10$ & $0.57\pm0.01$ & $0.54\pm0.01$ & $1.75\pm0.04$ &  $2.1\pm0.1$ & $3.0\pm0.1$ \\
 & \emph{z} & 0.44 & $1.13 \pm 0.14$ & $0.82\pm0.03$ & $0.86\pm0.01$  &  $1.69\pm0.09$ &  $2.3\pm0.3$ & $2.6\pm0.2$ \\
\cline{2-9}
2022 April 26 & \emph{g} & 1.25 & $2.42 \pm 0.08$ & $0.26\pm0.02$ & $0.36\pm0.02$ & $1.88\pm0.07$ & $6.7\pm0.4$ & $5.2\pm0.2$ \\
FNT-1 & \emph{r} & 1.02 & $2.10 \pm 0.09$ & $0.30\pm0.03$ & $0.38\pm0.02$ & $1.80\pm0.08$ & $7.8\pm0.7$ & $4.7\pm0.2$ \\
 & \emph{i} & 0.60 & $2.29 \pm 0.12$ & $0.39\pm0.07$ & $0.52\pm0.03$ & $1.68\pm0.08$ & $9.2\pm1.0$ & $5.3\pm0.2$\\
 & \emph{z} & 0.47 & $2.26 \pm 0.19$ & $0.60\pm0.14$ & $0.69\pm0.03$ & $1.80\pm0.12$ & $12.6\pm2.4$ & $6.2\pm0.7$ \\
 \cline{2-9}
2022 April 27 & \emph{g} & 1.28 & $1.30 \pm 0.08$ & $0.25\pm0.01$ & $0.36\pm0.03$ & $1.73\pm0.13$ & $3.8\pm0.6$ & $2.9\pm0.3$ \\
FNT-2 & \emph{r} & 1.04 & $1.03 \pm 0.09$ & $0.29\pm0.02$  & $0.38\pm0.02$ & $1.76\pm0.20$ & $5.0\pm1.4$ & $2.4\pm0.4$ \\
 & \emph{i} & 0.61 & $1.19 \pm 0.12$ & $0.38\pm0.06$  & $0.52\pm0.03$  &  $1.87\pm0.25$  &  $3.5\pm0.6$  & $2.2\pm0.2$  \\
 & \emph{z} & 0.47 & $1.03 \pm 0.16$ & $0.51\pm0.10$  & $0.63\pm0.04$ & $1.96\pm0.45$ & $3.0\pm0.6$ & $1.7\pm0.2$ \\
\enddata
\tablecomments{
Column (1): the dates when NGC 4395 was observed by GTC on 2018 April 16 \citep{McHardy2023MNRAS.519.3366M} and by FNT on 2022 April 26 and 27 \citep{Montano2022}. 
Column (2): the name of the observed band.
Column (3): $\bar f^x$ is the $x$-band mean flux density, where $x$ takes $g$, $r$, $i$, or $z$, in units of $\rm 10^{-15}~erg~s^{-1}~cm^{-2}~\text{\AA}^{-1}$.
Column (4): $\sigma_{\rm rms}$ is the excess variance, calculated using Equation~(8) of \citet{Vaughan2003MNRAS}, whose error is obtained using their Equation~(11).
Column (5): the mean and $1\sigma$ standard deviation of the $x$-band photometric uncertainties in magnitude, $\sigma^x_{\rm e}$, converted from the initial flux densities and the associated errors.
Columns (6) to (9): the best-fit parameters of $\sigma_n$, $\beta$, $\tau_{\rm d}$, and ${\rm SF}_\infty$ in Equation~(\ref{eq:fit}) as well as the corresponding $1\sigma$ uncertainties.
}
\end{deluxetable*}

\subsection{Structure Functions}\label{sec:sf}

The SF of a single-band light curve measures the typical variation amplitude as a function of time intervals (hereafter timescales). Generally, given an observed light curve, the observed SF can be derived by \citep[e.g.,][]{diClemente1996ApJ...463..466D,Kozlowski2016ApJ...826..118K}
\begin{equation}\label{eq:sf_obs}
    {\rm{SF}}_{\rm{obs}}(\uptau)=\sqrt{\frac{\pi}{2} \left \langle |m_i - m_j| \right \rangle^2}~~~~{\rm mag},
\end{equation}
where $m_i$ and $m_j$ are magnitudes at $t_i$ and $t_j$ instants, respectively, and the time interval between two instants $\uptau = |t_i - t_j|$. At short timescales, the observed SF is dominated by the photometric uncertainties, $\sigma_n$. Following the suggestion of \citet{Kozlowski2016ApJ...826..118K}, the underlying true SF, $\rm SF_{true}$, could be inferred through 
\begin{equation}\label{eq:fit}
   {\rm{SF}}^2_{\rm obs} (\uptau) = {{\rm SF}}^2_{\rm ture} (\uptau | \tau_{\rm d}, \beta, {\rm SF}_{\infty}) + 2\sigma_n^2
\end{equation}
once a specific functional form for the true SF has been assumed. For instance,
\begin{equation}\label{eq:sf_true}
    {{\rm SF}}_{\rm ture} (\uptau | \tau_{\rm d}, \beta, {\rm SF}_{\infty}) = {\rm SF}_{\infty} \sqrt{1- \exp \left( - \frac{\uptau}{\tau_{\rm d}} \right)^\beta },
\end{equation}
where $\tau_{\rm d}$ is the de-correlation or damping timescale, $\rm SF_\infty$ is the asymptotic rms variability on long timescales, and $\beta$ is the SF slope at $\uptau \ll \tau_{\rm d}$. Note that when $\uptau \rightarrow 0$, ${\rm SF_{obs}} \rightarrow \sqrt{2}\sigma_n$. 
On short timescales, i.e., $\ll \tau_{\rm d}$, the power-law slopes of the SF and PSD are $0.5 \beta$ and $- 2 \beta$, respectively.
Since \cite{Kelly2009ApJ...698..895K}, many works suggest that the DRW process with $\beta = 1$ can well describe the stochastic behavior of the optical AGN variability, in particular for quasars/AGNs with SMBHs over timescales of several weeks to years \citep[e.g.,][]{Kozlowshi2010ApJ...708..927K,MacLeod2010ApJ...721.1014M,Andrae2013A&A...554A.137A,Zu2013}.

\begin{figure*}[t!]
    \centering
    \includegraphics[width=0.45\textwidth]{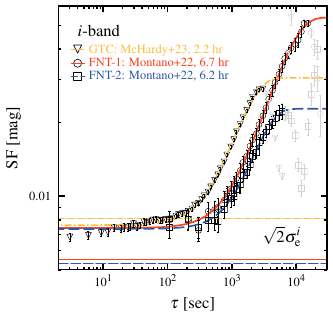}
    \includegraphics[width=0.45\textwidth]{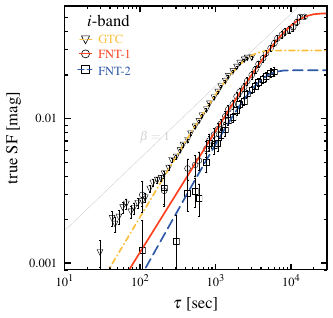}
    \caption{Left panel: The observed $i$-band FNT-1 (open circles), FNT-2 (open squares), and GTC (open triangles) SFs of NGC 4395 are fit by Equation~(\ref{eq:fit}) after excluding a few uncertain data points (light-gray) at long timescales (i.e., $\uptau > \uptau_{\rm max}$; see Section~\ref{sec:sf}). Including the best-fit photometric uncertainties, $\sigma_n$ (i.e., the horizontal part of the observed SFs at $\uptau \lesssim 10^{2-3}$~s; see Table~\ref{tab:fracs}), the best-fit FNT-1 (thick solid curve), FNT-2 (thick dashed curve), and GTC (thick dot-dashed curve) SFs are shown. 
    For comparison, the reported $i$-band mean photometric uncertainties, $\sigma^i_{\rm e}$, for the three datasets are shown in the same style but with thinner horizontal lines. Comparing the reported $\sigma^i_{\rm e}$ to the best-fit $\sigma_n$, \citet{McHardy2023MNRAS.519.3366M} may slightly overestimate the photometric uncertainties by a factor of $\simeq 1.1$, while \citet{Montano2022} may underestimate them by a factor of $\simeq 1.4$. Therefore, the best-fit $\sigma_n$ is used to determine the left borderline for the proper range of timescales of the SF (see Section~\ref{sec:timerange_SF}).
    Right panel: after subtracting the best-fit $\sigma_n$ (Equation~\ref{eq:fit}) and excluding unfit data points at $\uptau > \uptau_{\rm max}$, the corresponding true (symbols) and best-fit (curves) SFs are compared. The true SFs inferred from three nights have similar short-term slopes, $\beta \simeq 1.8$, regardless of the large differences in both $\tau_{\rm d}$ and ${\rm SF}_\infty$ (Table~\ref{tab:fracs}). Departures from the DRW process with $\beta = 1$ (the light-gray thin solid line) are prominent at timescales shorter than $\sim$hours in all three nights.
    }\label{fig:3nights-sf}
\end{figure*}

The left panel of Figure~\ref{fig:3nights-sf} illustrates the observed $i$-band SFs for NGC 4395 in three nights. Here, the observed SFs are measured between 1~s and $10^5$~s in steps of 0.05 dex, and the 1$\sigma$ uncertainties are estimated via the standard deviations of SFs derived from $10^3$ mock light curves, which are generated using the flux randomization/random subset selection method \citep[e.g.,][]{Fausnaugh2016ApJ...821...56F}. On timescales shorter than $\sim 10^{2-3}$~s, the observed SFs are flat because of the photometric uncertainties; on timescales longer than $\sim 10^4$~s, the observed SFs show large statistical fluctuation due to the limited baseline of the observed light curve.
The SFs of NGC 4395 varies between the three nights of observation, partly attributed to the randomness nature of AGN variability. 

Throughout this work we should present all results based on every night of observation on NGC 4395. However, we suggest that the results based on the FNT-1 light curves may be the most reliable.
On one hand, the FNT-1 light curves are the longest among the three nights available (Figure~\ref{fig:lc}). The FNT-1/2 light curves are longer by a factor of $\simeq 3$ than the GTC light curves. Although the FNT-1 light curves are merely longer by $\simeq 7.5\%$ than the FNT-2 light curves, the resultant FNT-1 SFs flatten at $\simeq 1.3 \times 10^4$~s, which is nearly $\simeq 2$ times the turnover timescale of the FNT-2 SFs (see the left panel of Figure~\ref{fig:3nights-sf} and the best-fit $\tau_{\rm d}$ in Table~\ref{tab:fracs}).
On the other hand, as demonstrated in Figure~\ref{fig:fvar}, both the variation amplitude ($\sigma_{\rm rms}$) and the significance of the variation ($\sigma_{\rm rms}/\sigma_{\rm e}$) of the FNT-1 light curves are larger by a factor of $\simeq 2$ than those of the FNT-2 and GTC light curves. 
Nevertheless, we would end up with our current conclusions considering also the results based on the FNT-2 light curves, which have comparable baseline to the FNT-1 light curves, and leaving firm conclusions for future more and better data on NGC 4395.

\subsubsection{A Proper Range of Timescales for the SF}\label{sec:timerange_SF}

Since the observed SFs at timescales longer than $\sim 10^4$~s are subject to substantial statistical fluctuations, we follow \citet{Emmanoulopoulos2010MNRAS.404..931E} and take a specific timescale, $\uptau_{\rm max}$, at the first local maximum of the observed SF as the right borderline for the proper range of timescales of the SF. 
In the left panel of Figure~\ref{fig:3nights-sf} the observed SFs beyond $\uptau_{\rm max}$ shown in the light-gray are not used in the following.
To determine a minimal timescale, $\uptau_{\rm min}$, as the left borderline for the proper range of timescales of the SF, we use the least squares method\footnote{\texttt{LsqFit.jl}: https://github.com/JuliaNLSolvers/LsqFit.jl} and fit Equation~(\ref{eq:fit}) to the observed SF after excluding the uncertain part beyond $\uptau_{\rm max}$. 

The left panel of Figure~\ref{fig:3nights-sf} illustrates the best-fit functions (three thick curves) for the three observed $i$-band SFs while the corresponding best-fit parameters (i.e., $\sigma_n$, $\beta$, $\tau_{\rm d}$, and ${\rm SF}_\infty$) for all bands are tabulated in Table~\ref{tab:fracs}. The best-fit $\sigma_n$ and $\beta$ are relatively reliable with small differences among three nights (cf. Columns (6) and (7) in Table~\ref{tab:fracs}). Instead, both $\tau_{\rm d}$ and ${\rm SF}_\infty$, even with quite small best-fit errors, are actually very uncertain owing to both the limited baseline and the randomness nature of AGN variability, also indicated by the large dispersion of the best-fit values among three nights (cf. Columns (8) and (9) in Table~\ref{tab:fracs}).

Comparing the reported photometric uncertainties to our best-fit $\sigma_n$ (the left panel of Figure~\ref{fig:3nights-sf} and Table~\ref{tab:fracs}), we find that \citet{McHardy2023MNRAS.519.3366M} may slightly overestimate the photometric uncertainties, while \citet{Montano2022} may somewhat underestimate them. Therefore, we decide to use the best-fit $\sigma_n$ in each band to determine $\uptau_{\rm min}$ for the corresponding band. Practically, we select $\uptau_{\rm min}$ such that ${\rm SF_{obs}}(\uptau_{\rm min}) = 2 \times \sqrt{2} \sigma_n$. Note that selecting ${\rm SF_{obs}}(\uptau_{\rm min}) = \sqrt{x} \times \sqrt{2} \sigma_n$, the noise contribution to the true SF at $\uptau_{\rm min}$ reaches a fraction of $\sqrt{x/(x-1)} - 1$, i.e., $\simeq 41\%$, $\simeq 10\%$, $\simeq 6\%$ for $x = 2$, 6, and 9, respectively.

Although we deem that at timescales between $\uptau_{\rm min}$ and $\uptau_{\rm max}$ the SF should have been measured fairly accurately, a timescale range narrower than $[\uptau_{\rm min}, \uptau_{\rm max}]$ suggested here is required by a more conservative consideration because (1) the SF with a smaller $\uptau_{\rm max}$ would be less affected by the limited baseline and (2) the SF with a larger $\uptau_{\rm min}$ would further decrease the noise contribution at $\uptau_{\rm min}$, which is still as high as $\simeq 15\% (= {\rm SF}_{\rm obs} / {\rm SF}_{\rm true} - 1)$ given our selection of ${\rm SF_{obs}}(\uptau_{\rm min}) = 2 \times \sqrt{2} \sigma_n$. 

In the right panel of Figure~\ref{fig:3nights-sf}, we show the so-called true SFs (symbols) and the corresponding best-fit ones (curves) after subtracting the best-fit noise term, $\sigma_n$, from the observed SFs and the corresponding best-fit ones, respectively.
A few unfit data points at $\uptau > \uptau_{\rm max}$ that are subject to the limited baseline are excluded. Comparing to the DRW prediction with $\beta = 1$, the true SFs in all three nights (and in all four bands; Table~\ref{tab:fracs}) have similar short-term slopes, $\beta \simeq 1.8$, demonstrating prominent departures from the DRW process at timescales shorter than $\sim$hours. Note that the residual noise at $\uptau \lesssim 200~{\rm s}$ still induces large fluctuations in the true SF and even a systematical offset from the best-fit one. Hereafter, the true SF always indicates the observed one subtracted the best-fit noise term.

\subsection{Color Variations}\label{sec:cv}

\begin{figure*}[t!]
    \centering
    \includegraphics[width=0.9\textwidth]{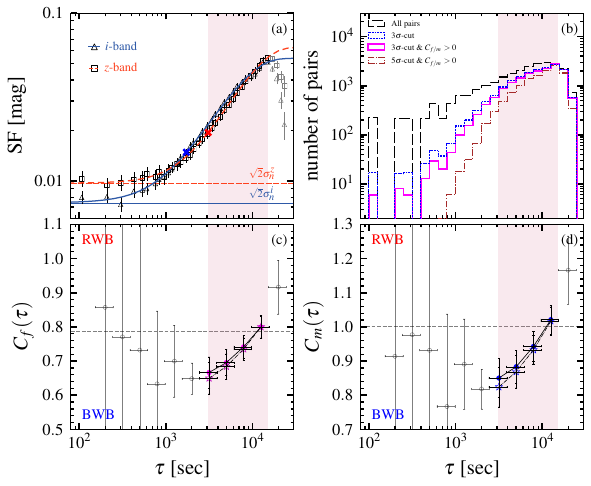}

    \label{fig: cv_sf}
    \caption{
    Using the $i$- and $z$-band FNT-1 light curves of NGC 4395, illustrated are the corresponding observed SFs (panel (a); Section~\ref{sec:sf}), distributions of data pairs (panel (b); Section~\ref{sec:cv}), and CVs in the flux/magnitude space (panel (c/d); Section~\ref{sec:cv}).
    Similar to Figure~\ref{fig:3nights-sf}, the panel (a) displays both the observed $i$- and $z$-band SFs (open symbols), the noise levels in terms of the best-fit $\sqrt{2}\sigma_n$ (two horizontal lines), the minimal timescales where ${\rm SF_{obs}} = 2 \times \sqrt{2}\sigma_n$ adopted for each band (cross symbols), and a shaded region for the proper range of timescales, deemed to be little affected by both the photometric uncertainties and the limited baseline (Section~\ref{sec:timerange_CV}).
    The panel (b) shows distributions of data pairs: all pairs (dashed histogram), pairs after $3\sigma$-cut (dotted histogram), pairs satisfying both $3\sigma$-cut and $C_{f/m} > 0$ (solid histogram), and pairs satisfying both $5\sigma$-cut and $C_{f/m} > 0$ (dot-dashed histogram).
    The panel (c/d) presents, using pairs satisfying both $3\sigma$-cut and $C_{f/m} > 0$, the CVs in the flux/magnitude space (circles), while highlighted are a subset possessing a definite timescale-dependent within the shaded region (blue filled circles). 
    For comparison, shown are the CVs (open asterisks) implied by pairs satisfying both $5\sigma$-cut and $C_{f/m} > 0$. Besides, two horizontal dashed lines in the two bottom panels separate the CVs into the BWB and RWB regions.
}\label{fig:i-z}
\end{figure*}

The SF provides the timescale-dependent variation property in a single band, while the CV highlights more subtle timescale-dependent variation property between two bands.
In terms of a so-called quantity $\theta$, defined in \citet{sun2014ApJ...792...54S}, the timescale-dependent CV was originally presented in the magnitude space by \citet[][see also \citealt{Zhu2016ApJ...832...75Z}]{sun2014ApJ...792...54S} and also in the flux space by \citet{Cai2016}. 
Later on, simplifying the analysis of the CV was upgraded by \citet{Zhu2018} and \citet{Cai2019} in the flux space and the magnitude space for local Seyfert galaxies and quasars, respectively.

It is worth noting that analyzing the CV in the magnitude space is highly immune to the intrinsic host extinction but can be subject to the contamination of the host emission and thus is more applicable to quasars whose host emission is negligible. 
However, for Seyfert galaxies and low-luminosity AGNs, such as NGC 4395, with significant host contribution, analyzing the CV in the flux space can naturally remove the host component, though the resulting color can be somewhat (timescale-independent) offset by the extinction difference between two given bands.

Here, we briefly outline the basic procedure for the CV in both the flux and magnitude spaces ($C_{f/m}$). Figure~\ref{fig:i-z} exemplifies the CV using the $i$- and $z$-band FNT-1 light curves because (1) these two bands suffer less from the contamination of broad emission lines (Figure \ref{fig:trans_spec}) and (2) the resultant CV is the most significant (Section~\ref{sec:comp_cv}).

Given two arbitrary light curves with superscripts ``$r$'' and ``$b$'' representing the redder and bluer bands, respectively, 

(i) re-bin the two light curves such that they have photometric measurements ($f_i$ or $m_i$) and associated uncertainties ($\sigma_i$) at the same instants ($t_i$), as we discussed in Section~\ref{sec:lc};

(ii) calculate the flux (magnitude) differences between the same two instants of two bands, e.g., $\Delta f^r(\uptau) = f_i^r - f_j^r$ and $\Delta f^b(\uptau) = f_i^b - f_j^b$ for the same $\uptau = |t_i - t_j|$, forming $C^2_n$ data pairs, \{..., [$\Delta f^b(\uptau)$, $\Delta f^r(\uptau)$], ...\};

(iii) perform a so-called $3\sigma$-cut to exclude data pairs with statistically insignificant variations, i.e., requiring $\sqrt{(\Delta f^b)^2 + (\Delta f^r)^2} > 3 \times \sqrt{\Sigma_i^2 + \Sigma_j^2}$ where a global photometric uncertainty at the $i/j$ instant, being composed of the photometric uncertainties in the two bands, is given by $\Sigma_{i/j} = \sigma^b_{i/j} \sigma^r_{i/j} / \sqrt{(\sigma^b_{i/j} \sin \phi)^2 + (\sigma^r_{i/j} \cos \phi)^2}$ with $\tan \phi \equiv \Delta f^r/\Delta f^b$ (cf. \citealt{sun2014ApJ...792...54S} and their Figure~1 for details). 
Note that applying the $3\sigma$-cut primarily decreases the number of data pairs at shorter timescales (i.e., from the black dashed histogram to the blue dotted histogram in the panel (b) of Figure \ref{fig:i-z});

(iv) assess the CV according to, in the flux space,
\begin{equation}\label{eq:color_flux}
    C_f(\uptau)=\frac{\Delta f^r(\uptau)}{\Delta f^b(\uptau)}=\frac{f^r_i-f^r_j}{f^b_i-f^b_j},
\end{equation}
where $f_i$ is in units of $\rm erg/s/cm^{2}/\text{\AA}$, or in the magnitude space,
\begin{equation}\label{eq:color_mag}
    C_m(\uptau)=\frac{\Delta m^r(\uptau)}{\Delta m^b(\uptau)}=\frac{m^r_i-m^r_j}{m^b_i-m^b_j},
\end{equation}
shown in the panels (c) and (d) of Figure \ref{fig:i-z}, respectively;

(v) take the median value of a set of $C_{f/m}$ within a certain range of timescales after excluding negative or unphysical $C_{f/m}$, that is, requiring that AGNs behave brighter or fainter simultaneously in both bands and avoiding bias by the extreme $C_{f/m}$. Note that requiring $C_{f/m} > 0$ only removes a few data pairs, again mostly at shorter timescales (i.e., from the blue dotted histogram to the magenta solid histogram in the panel (b) of Figure \ref{fig:i-z});

and (vi) utilize the random subset selection method following \cite{Zhu2018} to estimate the uncertainty for the median CV derived in the step (v), that is, generating $10^3$ realizations of mock light curves (with replacement), repeating steps (i) to (v), and taking the standard deviation of the bootstrapped median CV as the desired 1$\sigma$ uncertainty.

The final data pairs satisfying both the $3\sigma$-cut and $C_{f/m} > 0$ are used to calculate the CV as a function of timescales in the flux and magnitude space, shown as the open/filled circles in the panels (c) and (d) of Figure \ref{fig:i-z}, respectively. Since the CV can be approximated by the ratio of two SFs and so fluctuates more strongly than the SF, a larger bin size of 0.2 dex is adopted to average $C_{f/m}$. In panels (c) and (d) of Figure \ref{fig:i-z}, there are two horizontal dashed lines, beneath and beyond which the CV behaves BWB and redder-when-bright (RWB), respectively.
In the magnitude space, $C_{m} = 1$ is always the separation between BWB and RWB. Instead, in the flux space, the same separation depends on the relative contribution between two given bands and can be roughly estimated as $C_f \simeq ({\bar f^r}/{\bar f^b}) C_m = {\bar f^r}/{\bar f^b}$ where $\bar f^r$ and $\bar f^b$ are the mean flux densities in the redder and bluer bands (see Equation~(4) of \citealt{sun2014ApJ...792...54S}).

For $3 \times 10^3~{\rm s} \lesssim \uptau \lesssim 10^4~{\rm s}$, the $i$ versus $z$ CVs of NGC 4395 in both the flux and magnitude spaces all behave BWB and are clearly timescale-dependent, that is, the BWB trend is more prominent at shorter timescales. 
Instead, once $\uptau \lesssim 3 \times 10^3$~s, the BWB trend gradually disappears with decreasing timescales. This is mainly attributed to the increasing dominance of the photometric uncertainty with decreasing timescales as simulated by \citet[][cf. their Figure 5]{Zhu2018}.
On the other aspect of $\uptau \gtrsim 10^4$~s, the CV seemingly behaves RWB and becomes more redder at longer timescales. Although some mechanisms, such as the inward propagation of the perturbation over the accretion disk, could result in a RWB trend at longer timescales, this RWB trend at $\uptau \gtrsim 10^4$~s is not conclusive because it could be heavily subject to the limited baseline as indicated by the rapid drops of both the SF and the number of data pairs, shown in the panel (a) and (b) of Figure~\ref{fig:i-z}, respectively.

\subsubsection{A Proper Range of Timescales for the CV}\label{sec:timerange_CV}

As introduced above, the CVs over all available timescales are globally complicated and definitely affected by the photometric uncertainty and the limited baseline at shorter and longer timescales, respectively. We therefore seek a unique way to select a proper range of timescales in which the derived CVs are less affected by both the photometric uncertainty and the limited baseline. 

We rely on the observed SF for which we have determined a proper range of timescales in a single band (Section~\ref{sec:sf}). For instance, assuming $[\uptau_{\rm min}^i, \uptau_{\rm max}^i]$ and $[\uptau_{\rm min}^z, \uptau_{\rm max}^z]$ are the proper ranges of timescales for the observed $i$- and $z$-band SFs, respectively, we thus treat $[\max\{\uptau_{\rm min}^i, \uptau_{\rm min}^z\}, \min\{\uptau_{\rm max}^i, \uptau_{\rm max}^z\}]$ as the proper range of timescales (cf. the shaded regions in Figure~\ref{fig:i-z}) for the $i$ versus $z$ CV.

By increasing the $3\sigma$-cut to a stricter $5\sigma$-cut, we confirm that the resultant CVs within the proper range of timescales are also consistent (Figure~\ref{fig:i-z}), and thus we adopt the $3\sigma$-cut in the following discussions.

\section{Discussions}\label{sec:dis}

\begin{figure*}[t!]
    \centering
    \includegraphics[width=0.6\textwidth]{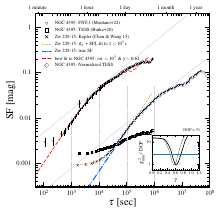}
    \includegraphics[width=0.4\textwidth]{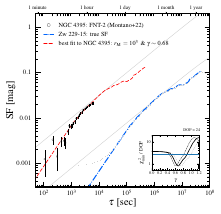}
    \includegraphics[width=0.4\textwidth]{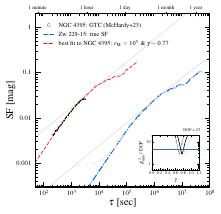}
    \caption{Top panel: SFs of NGC 4395 ($M_{\rm BH} \simeq 10^4~M_\odot$) at the intermediate-mass scale are compared to that of Zw 229-15 ($M_{\rm BH} \simeq 10^7~M_\odot$) at the supermassive scale. 
    For NGC 4395, the true $i$-band FNT-1 SF \citep[][circles]{Montano2022} is shown with the observed $I_{\rm C}$-band TESS SF \citep[][squares]{Burke2020ApJ...899..136B} as well as the corresponding normalized ``true'' TESS SF (diamonds).
    For Zw 229-15, its {\it Kepler} light curve stitched by \citet[][]{chen2015} is used to derive the observed SF (crosses), a short-term portion (i.e., $\uptau < 10^5$~s) of which is fitted with a noise term ($\sigma_n$) plus a single power-law (SPL) function (i.e., the dotted curve for the best $\sigma_n$ + SPL fit with $\beta \simeq 1.82 \pm 0.01$, that has been extended beyond $\uptau \sim 10^5$~s to highlight the complexity of the true SF more than a single SPL). After removing the noise contribution, the true SF of Zw 229-15 (the dot-dashed curve) is scaled horizontally by a factor of $r_{\rm M}^\gamma$ before fitted vertically to the true FNT-1 SF of NGC 4395, where $r_{\rm M}$ is the mass ratio of Zw 229-15 to NGC 4395 and $\gamma$ is the mass index.
    The inset shows the minimal chi-square, $\chi^2_{\rm min}$, divided by the degree of freedom (DOF), as a function of $\gamma$ for $r_{\rm M} = 10^3$ (solid curve), $r_{\rm M}/2$ (dotted curve), and $2 r_{\rm M}$ (dashed curve). A blue horizontal line in the inset represents the 3$\sigma$ confidence level given the DOF.
    The best-fit SF to NGC 4395 (the dashed curve) is presented for $r_{\rm M} = 10^3$ and the optimized $\gamma \sim 0.62$. 
    Interestingly, the true FNT-1 and ``true'' TESS SFs of NGC 4395 across $\sim 10^3$~s to $\sim 10^6$~s is found to be in nice agreement with the scaled SF of Zw 229-15, suggesting a similar accretion physics in AGNs across different mass scales.
    For comparison, two light-gray thin lines tilted for $\beta = 1$ highlight prominent departures from the DRW process at short timescales.
    Bottom panels: same as the top panel but for fitting the true SF of Zw 229-15 to the FNT-2 (bottom-left panel) and GTC (bottom-right panel) SFs of NGC 4395. The fact that somewhat larger $\gamma$ is preferred by a shorter baseline indicates a potential bias on $\gamma$ induced by the shorter baseline.
    }\label{fig:sf_dis}
\end{figure*}

\subsection{Structure Function (SF)}\label{sec:sf_dis}

\subsubsection{Prominent departures from the DRW process at the intermediate-mass scale}\label{sec:sf_4395}

Although departures from the DRW process have been reported in AGNs at the supermassive scale \citep[e.g.,][]{Mushotzky2011ApJ...743L..12M,Kelly2014,Kasliwal2015,Simm2016A&A...585A.129S,Zhu2016ApJ...832...75Z,Guo2017ApJ...847..132G,stone2023}, a departure alike has never been reported in any AGN at the intermediate-mass scale. Thanks to the high-quality (i.e., high cadence, long baseline, small photometric uncertainties) light curves provided by current surveys, we now have the chance to study the variability properties of an IMBH, NGC 4395, at timescales of minutes to hours \citep{Burke2020ApJ...899..136B, McHardy2023MNRAS.519.3366M, Montano2022}.

As tabulated in Table \ref{tab:fracs} for all bands and nights, all values of the best-fit $\beta$ for NGC 4395 indicate significant departures from the DRW process at the intermediate-mass scale for the first time.
For the $g$, $r$, $i$, and $z$ bands, the mean $\beta$ over three nights are $\simeq 1.80 \pm 0.05$, $\simeq 1.80 \pm 0.07$, $\simeq 1.77 \pm 0.09$, and $\simeq 1.82 \pm 0.16$, respectively, which are all inconsistent with $\beta\equiv1$ implied by the DRW process.
After subtracting the best-fit $\sigma_n$ and excluding unfit data points, the right panel of Figure~\ref{fig:3nights-sf} illustrates the $i$-band true SFs of NGC 4395 observed in three nights. Regardless of the large differences in both $\tau_{\rm d}$ and ${\rm SF}_\infty$ between the three nights of observation, owing to both the limited baseline and the randomness nature of AGN variability, almost the same short-term slopes at timescales shorter than $\sim$hours are found in those SFs of NGC 4395 and are all steeper than implied by the DRW process.

Previously, \cite{Burke2020ApJ...899..136B} analyzed the $I_{\rm C}$-band\footnote{The TESS bandpass, i.e., 6000~\angstrom~to 1~$\mu$m (\citealt{TESS}), is much wider than but is centered on the Johnson-Cousins $I_{\rm C}$ band, whose effective wavelength is 7980~\angstrom~(\citealt{Bessell2005ARA&A..43..293B}). Therefore, we compare the $i$- and $I_{\rm C}$-band SFs.} TESS light curve ($\sim$month-long and 30-minute-cadence) of NGC 4395 and found $\beta \simeq 0.94 \pm 0.08$, which is consistent with the DRW process. Using their light curve, the top panel of Figure~\ref{fig:sf_dis} shows the observed TESS SF (squares) which can be fitted by Equation~(\ref{eq:fit}) with $\beta \simeq 0.96 \pm 0.05$. 
Since the TESS SF is limited to timescales $\gtrsim 2 \times 10^3$~s and strongly affected by photometric uncertainties at timescales $\lesssim 10^4$~s, the $\beta$ derived from the TESS SF should have only characterized the slope of SF around $\sim 10^4$~s and can not represent the true $\beta$ at $\lesssim 10^4$~s, while we can using the much higher cadence and more accurate light curves from \citet{Montano2022} and \citet{McHardy2023MNRAS.519.3366M}.

In the top panel of Figure~\ref{fig:sf_dis}, the observed TESS SF (squares) is significantly lower than the FNT-1 SF (circles), probably owing to the large TESS pixels (21 arcsec), which results in huge galaxy contamination.
Here, after subtracting the best-fit $\sigma_n$, scaling the ``true'' TESS SF horizontally to the same level as the FNT-1 SF at $\sim 10^4$~s (diamonds) forms a stitched global SF spanning $\sim$minute to $\sim$month for NGC 4395. This stitched SF suggests that its slope may change from $\beta \simeq 1.8$ below $\sim 10^4$~s, through $\beta \simeq 1.0$ around, and to $\beta \simeq 0.7$ beyond, to be justified by future homogeneous observations on NGC 4395. Even quite uncertain, this comparison between the TESS and FNT-1 SFs of NGC 4395 is suggestive and could stimulate more future observations on it spanning timescales over $\sim$minute to $\sim$month, and even $\sim$year.

To obtain a true global SF for NGC 4395, longer (as long as several months) and denser (as short as $\sim 100$~s) monitoring should be conducted repeatedly.
Using $\sim$month long light curve of NGC 4395, \citet{Burke2020ApJ...899..136B} find a damping timescale of $1.4^{+1.9}_{-0.5}$ days. The $1\sigma$ upper limit for the damping timescale is already 3.3 days, suggesting that a single one-month-long light curve is far from any firm conclusion on the true intrinsic damping timescale for NGC 4395 (e.g., \citealt{Hu2023arXiv231016223H, zhou2024}). According to \citet[][cf. their Figure~8]{Hu2023arXiv231016223H}, if the true damping timescale is 1.4 days for NGC 4395, the retrieved damping timescale has $\sim 40\%$ ($1\sigma$ confidence level) dispersion given a single light curve with a one-month-long baseline. For NGC 4395, the dispersion can be reduced to $\sim 10\%$ ($1\sigma$ confidence level) only if the baseline of a continuing monitoring has been increased to $\sim $year or a one-month-long monitoring has been repeated at least 10 times.

\begin{deluxetable*}{ccc ccc ccc}\label{tab:smbh}
\tablewidth{0pt}
\tablecaption{Basic physical and observational properties of four AGNs at the supermassive scale.}
\tablehead{\colhead{AGN Name} & \colhead{$M_{\rm BH}$} & \colhead{$\lambda_{\rm Edd}$} & \colhead{$\Delta t_{\rm min}$} & \colhead{$\Delta t_{\rm max}$} & \colhead{$\sigma_n$} & \colhead{$\beta_1$} & \colhead{$\beta_2$} & \colhead{$\beta_3$} \\
\cline{7-9}
\colhead{} & \colhead{($10^7 M_\sun$)} & \colhead{} & \colhead{} & \colhead{(yr)} & \colhead{($10^{-2}$ mag)} & \colhead{[$\uptau_1$, $\uptau_2$]/(s)} & \colhead{[$\uptau_2$, $\uptau_3$]/(s)} & \colhead{[$\uptau_3$, $\uptau_4$]/(s)} \\
\colhead{(1)} & \colhead{(2)} & \colhead{(3)} & \colhead{(4)} & \colhead{(5)} & \colhead{(6)} & \colhead{(7)} & \colhead{(8)} & \colhead{(9)}
}
\startdata
Zw 229-15 & $\simeq 1$ & 0.050 & $30$ min & $\simeq 3.30$ & $0.07 \pm 0.01$ & $\simeq 1.82 \pm 0.01$ & $\simeq 1.27 \pm 0.01$ & $\simeq 0.73 \pm 0.01$ \\
 & & & & & & [$10^3, 10^5$] & [$10^5, 10^6$] & [$10^6, 5\times10^7$] \\
\cline{7-9}
W2 1925+50 & $\simeq 10$ & 0.013 & $30$ min & $\simeq 2.79$ & $0.11 \pm 0.01$ & $\simeq 1.83 \pm 0.01$ & $\simeq 1.30 \pm 0.01$ & $\simeq 0.70 \pm 0.01$ \\
 & & & & & & [$10^3, 5\times10^5$] & [$5\times10^5, 10^7$] & [$10^7,  5\times10^7$] \\
\cline{7-9}
W2R 1904+37 & $\simeq 5$ & 0.089 & $30$ min & $\simeq 2.54$ & $0.10 \pm 0.01$ & $\simeq 1.83 \pm 0.01$ & - & - \\
& & & & & & [$10^3, 10^6$] & - & - \\
\cline{7-9}
NGC 5548 & $\simeq 5$ & 0.050 & $\simeq 1$ day & $\simeq 0.58$ & $0.36 \pm 0.10 $& $\simeq 1.72 \pm 0.16$ & - & - \\
& & & & & & [$10^4, 5\times10^5$] & - & - \\
\enddata
\tablecomments{
Column (1): the names of AGNs at the supermassive scale. 
Column (2): the BH mass in units of $10^7~M_\odot$, taken from \cite{Barth2011ApJ...732..121B} for Zw 229-15, from \cite{Smith2018ApJ...857..141S} for W2 1925+50 and W2R 1904+37, and from \cite{Fausnaugh2016ApJ...821...56F} for NGC 5548. 
Column (3): the Eddington ratio taken from the same references as the BH mass. 
Column (4): the minimal cadence given by the {\it Kepler} mission for the first three AGNs, while the median one for NGC 5548 taken from \cite{Fausnaugh2016ApJ...821...56F}.
Column (5): the maximal available baseline in units of year, estimated using the stitched {\it Kepler} light curves from \cite{chen2015} and the $i$-band one from \cite{Fausnaugh2016ApJ...821...56F} for NGC 5548. 
Column (6): the best-fit noise level.
Columns (7) to (9): the best-fit SPL slopes as well as the corresponding 1$\sigma$ uncertainties within three timescale ranges given by $[\uptau_i, \uptau_{i+1}]$ in units of second. For the smallest timescale range, $[\uptau_1, \uptau_2]$, a noise level presented in Column (6) has been included in the fit, while a single SPL function is used for the other two timescale ranges.
}
\end{deluxetable*}

\begin{deluxetable*}{ccc ccc ccc}\label{tab:mass_gamma}
\tablewidth{0pt}
\tablecaption{Comparing the true SFs of four AGNs at the supermassive scale to the true SF of NGC 4395 observed in three nights at the intermediate-mass scale to determine the best $\gamma$ for $\uptau \propto M^\gamma$.}
\tablehead{
\colhead{AGN Name} & \colhead{$r_{\rm M}$} & \colhead{NGC 4395} & \colhead{DOF} & \colhead{$\Delta \chi^2$(DOF)} & \colhead{Min[$\chi^2_{\rm min}$/DOF]} & \colhead{} & \colhead{best $\gamma$} \\
\cline{7-9}
& & & & & & \colhead{$r_{\rm M}/2$} & \colhead{$r_{\rm M}$} & \colhead{$2 r_{\rm M}$} \\
\colhead{(1)} & \colhead{(2)} & \colhead{(3)} & \colhead{(4)} & \colhead{(5)} & \colhead{(6)} & \colhead{(7)} & \colhead{(8)} & \colhead{(9)}
}
\startdata
& & FNT-1 & 31 & 57.40 & 0.88 & $\sim 0.69^{+0.13}_{-0.12}$ & \pmb{$\sim 0.62^{+0.11}_{-0.10}$} & $\sim 0.56^{+0.11}_{-0.10}$ \\
Zw 229-15 & $10^3$ & FNT-2 & 24 &  47.76 & 0.79 & $\sim 0.76^{+0.22}_{-0.20}$  & \pmb{$\sim 0.68^{+0.22}_{-0.20}$} & $\sim 0.62^{+0.20}_{-0.18}$ \\
(Figure~\ref{fig:sf_dis}) & & GTC & 23 & 46.36 & 2.70 & $\sim 0.86^{+0.04}_{-0.03}$  & $\sim 0.77^{+0.03}_{-0.03}$ & $\sim 0.70^{+0.03}_{-0.03}$ \\
\cline{1-9}
 & & FNT-1 & 31 & 57.40 & 0.54 & $\sim 0.78^{+0.05}_{-0.07}$ & \pmb{$\sim 0.72^{+0.11}_{-0.06}$} & $\sim 0.67^{+0.05}_{-0.06}$ \\
W2 1925+50 & $10^4$ & FNT-2 & 24 & 47.76 & 0.87 & $\sim 0.86^{+0.09}_{-0.30}$ & \pmb{$\sim 0.80^{+0.08}_{-0.28}$} & $\sim 0.74^{+0.08}_{-0.26}$\\
(Figure~\ref{fig:sf_dis_more_kepler2}) & & GTC & 23 & 46.36 & 2.84 & $\sim 0.93^{+0.02}_{-0.05}$ & $\sim 0.86^{+0.02}_{-0.05}$ & $\sim 0.80^{+0.02}_{-0.04}$ \\
\cline{1-9}
 & & FNT-1 & 31 & 57.40 & 1.10 & $\sim 0.92^{+0.03}_{-0.03}$ & $\sim 0.85^{+0.01}_{-0.03}$ & $\sim 0.78^{+0.02}_{-0.02}$ \\
W2R 1904+37 & $5 \times 10^3$ & FNT-2 & 24 & 47.76 & 0.58 & $\sim 1.03^{+0.04}_{-0.37}$ & $\sim 0.95^{+0.03}_{-0.33}$ & $\sim 0.87^{+0.04}_{-0.03}$ \\
(Figure~\ref{fig:sf_dis_more_kepler3}) & & GTC & 23 & 46.36 & 1.70 & $\sim 1.12^{+0.01}_{-0.02}$ & $\sim1.03^{+0.01}_{-0.02}$ & $\sim 0.95^{+0.01}_{-0.01}$ \\
\cline{1-9}
 & & FNT-1 & 31 & 57.40 & 0.43 & $\sim0.65^{+0.04}_{-0.05}$ & $\sim 0.59^{+0.04}_{-0.04}$ & $\sim 0.55^{+0.03}_{-0.04}$ \\
NGC 5548 & $5 \times 10^3$ & FNT-2 & 24 & 47.76 & 0.67 & $\sim 0.75^{+0.07}_{-0.75}$ & $\sim 0.68^{+0.07}_{-0.68}$ & $\sim 0.63^{+0.06}_{-0.63}$ \\
(Figure~\ref{fig:sf_dis_more_5548}) & & GTC & 23 & 46.36 & 0.92 & $\sim 0.83^{+0.02}_{-0.03}$ & $\sim 0.76^{+0.02}_{-0.02}$ & $\sim 0.70^{+0.02}_{-0.02}$ \\
\enddata
\tablecomments{
Column (1): the names of AGNs at the supermassive scale. 
Column (2): the fiducial mass ratios of AGNs at the supermassive scale to that of NGC 4395 ($M_{\rm BH} \simeq 10^4~M_\odot$).
Column (3): the abbreviations indicate observations on NGC 4395 in three nights. 
Column (4): the degree of freedom (DOF) in each fit.
Column (5): the expected change in $\chi^2$ given a $3\sigma$ confidence level, estimated following \citet[][cf. their Section~5.3.1]{Wall2012}.
Column (6): the minimum of $\chi^2_{\rm min}(\gamma)$/DOF, where a minimal $\chi^2_{\rm min}$ is determined for each given $r_{\rm M}$ and $\gamma$. Changing $r_{\rm M}$ results in distinct best $\gamma$, but the corresponding value of Min[$\chi^2_{\rm min}(\gamma)$/DOF] is the same.
Columns (7) to (9): the best values for $\gamma$ corresponding to the Min[$\chi^2_{\rm min}(\gamma)$/DOF], as well as the associated $3\sigma$ uncertainties, given three fixed values for the mass ratio, i.e., $r_{\rm M}$, $r_{\rm M}/2$, and $2 r_{\rm M}$.
Highlighted in boldface are the most plausible values for $\gamma$ we suggest according to the available data: both the {\it Kepler} light curves for Zw 229-15 and W2 1925+50 as well as the FNT-1/2 light curve for NGC 4395 are all or nearly the longest.
}
\end{deluxetable*}

\subsubsection{SFs of AGNs at the supermassive scale}\label{sec:sf_smbh}

The optical variability properties of individual AGNs \citep[e.g.,][]{Mushotzky2011ApJ...743L..12M,Kasliwal2015,Simm2016A&A...585A.129S,Zhu2018} and AGN samples \citep[e.g.,][]{Arevalo2023, arevalo2023a, stone2023} at the supermassive scale have been extensively investigated in recent years. However, there are only a few AGNs possessing both sufficiently long (as long as several years) and high cadence (as short as $\sim$hour) monitoring such that the global variation properties across several orders of timescales can be unveiled, rather than a single typical damping timescale ($\tau_{\rm d}$), which is also uncertain owing to the limited baseline and cadence. 
Instead, we would directly compare the global SFs of several AGNs, mainly from the {\it Kepler} monitoring, at the supermassive scale to that of NGC 4395 at the intermediate-mass scale in order to investigate the accretion physics across different mass scales.

Benefiting from the sufficiently long ($\sim$years) and high cadence ($\sim 30$ minutes) {\it Kepler} monitoring in a broad bandpass from 4230~\angstrom~to 8970~\angstrom~\citep{Koch2010}, \cite{Mushotzky2011ApJ...743L..12M} first reported departures from the DRW process at timescales of $\sim 2 \times 10^4$~s (or $\sim$ 6~hr) to $\sim 2 \times 10^6$~s (or $\sim 1$ month) in four AGNs at the supermassive scale. Among them, the Seyfert galaxy Zw 229-15 (at $z = 0.0275$ and with $M_{\rm BH} \simeq 10^7~M_\odot$; \citealt{Barth2011ApJ...732..121B}) has been observed in four quarters, within which $\beta$ changes from $\simeq 1.48$ to $\simeq 1.65$; 
the Seyfert galaxy W2R 1904+37 (at $z = 0.089$ with $M_{\rm BH} \simeq 10^{7.66}~M_\odot$; \citealt{Smith2018ApJ...857..141S}) has been observed in three quarters, within which changes from $\beta$ $\simeq 1.30$ to $\simeq 1.38$; 
the Seyfert galaxy W2 1925+50 (at $z = 0.067$ with $M_{\rm BH} \simeq 10^{8.04}~M_\odot$; \citealt{Smith2018ApJ...857..141S}) has been observed in three quarters, within which $\beta$ changes from $\simeq 1.37$ to $\simeq 1.48$; 
and the Seyfert galaxy W2R 1858+48 has been observed in two quarters, within which $\beta$ transitioned from $\simeq 1.44$ to $\simeq 1.49$.
Except the last one, the other three AGNs, i.e., Zw 229-15, W2R 1904+37, and W2 1925+50, have years-long light curves stitched by \cite{chen2015}\footnote{The light curves of W2R 1858+48 can not be stitched according to \cite{chen2015}. Instead, they stitched the light curves of CGRaBS J1918+4937. However, we do not consider CGRaBS J1918+4937 since it is likely a BL Lac object without a reliable BH mass \citep{Smith2018ApJ...857..141S}.} such that their global SFs are achievable.
Therefore, we mainly consider these three AGNs at the supermassive scale. 
One more AGN at the supermassive scale we consider even without sufficiently long and high cadence monitoring (with $\sim 200$ days-long and $\sim 1$ day-cadence according to \citealt{Fausnaugh2016ApJ...821...56F}) is the famous Seyfert galaxy NGC 5548 (at $z = 0.017175$ with $M_{\rm BH} \simeq 10^{7.59}~M_\odot$; \citealt{Pancoast2014MNRAS.445.3073P}) since it is the first individual AGN whose CVs between UV/optical are found to be timescale-dependent by \cite{Zhu2018} and are to be compared with that of NGC 4395 in Section~\ref{sec:comp_cv}. To compare with the $i$-band SFs of NGC 4395, we also use the $i$-band light curve for NGC 5548 from \citet{Fausnaugh2016ApJ...821...56F}.

Basic physical and observational properties are tabulated in Table~\ref{tab:smbh} for the four AGNs at the supermassive scale. In the top panels of Figures~\ref{fig:sf_dis} to \ref{fig:sf_dis_more_5548}, we show the observed SFs (crosses) for Zw 229-15, W2 1925+50, W2R 1904+37, and NGC 5548, respectively.
For the two AGNs with the longest baselines, i.e., Zw 229-15 over $\simeq 3.3$ yrs and W2 1925+50 over $\simeq 2.8$ yrs, their SFs intuitively have at least two breaks near $\sim 10^{5-6}$~s and $\sim 10^{6-7}$~s, indicating that a single $\beta$ could be not enough for describing their global SFs. Using a single power-law (SPL) function fitting to the SF in three distinct timescale ranges, we find that with increasing timescales $\beta$ changes from $\simeq 1.8$, through $\simeq 1.3$, to $\simeq 0.7$ for both Zw 229-15 and W2 1925+50 (see Table~\ref{tab:smbh} for the selected timescale ranges and the corresponding best-fit $\beta$). Note that owing to the significance of the noise contribution within the smallest timescale range, a noise term has been included in the fit, that is, the nominated $\sigma_n$ + SPL fit (cf. the dotted curves in Figures~\ref{fig:sf_dis} and \ref{fig:sf_dis_more_kepler2} for Zw 229-15 and W2 1925+50, respectively).

Similarly, we have performed the $\sigma_n$ + SPL fit to the observed SFs of the other two AGNs at the supermassive scale, i.e., W2R 1904+37 (Figure~\ref{fig:sf_dis_more_kepler3}) and NGC 5548 (Figure~\ref{fig:sf_dis_more_5548}). However, likely owing to their relatively short baselines, only a single short-term slope for each of them can be determined.

Interestingly, at timescales shorter than $\sim 10^{5-6}$~s, the short-term slopes of SFs of the four AGNs at the supermassive scale are all around $\beta \simeq 1.8$ (Table~\ref{tab:smbh}), which is also nicely consistent with that of NGC 4395 at $\uptau \lesssim 10^4$~s (Section~\ref{sec:sf_4395} and the right panel of Figure~\ref{fig:3nights-sf}). The consistency of the short-term slopes of the SFs among AGNs across a broad mass scale suggests a similar accretion physics at work.

The five AGNs we consider here show almost the same departures from the DRW process. \cite{Mushotzky2011ApJ...743L..12M} suggest the departure from the DRW process may be related to the magneto-rotational instability in terms of a similar high-frequency slope of PSDs of the mass accretion rate simulated by \citet{Reynolds2009ApJ...692..869R}.
Regardless of the fundamental problem suffered by simulations to convert the disk turbulence into radiation, we unveil more complicated structures in the global SFs of Zw 229-15 and W2 1925+50, suggesting that a global comparison between simulations and the data is necessary and essential.

\subsubsection{Comparing SFs of AGNs across the intermediate-mass to supermassive scales}\label{sec:comp_sf}

\begin{figure*}[t!]
    \centering
    \includegraphics[width=0.6\textwidth]{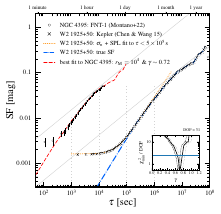}
    \includegraphics[width=0.4\textwidth]{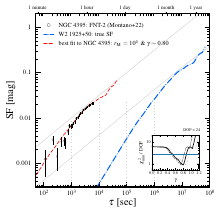}
    \includegraphics[width=0.4\textwidth]{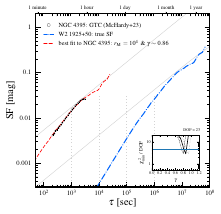} 
        \caption{Same as Figure~\ref{fig:sf_dis}, but for comparing NGC 4395 ($M_{\rm BH} \simeq 10^4~M_\odot$) with W2 1925+50 ($M_{\rm BH} \simeq 10^8~M_\odot$) and $r_{\rm M} = 10^4$. The observed SF of W2 1925+50 (crosses) is derived from the {\it Kepler} light curves stitched by \cite{chen2015}. The $\sigma_n$ + SPL fit is limited to $\uptau \lesssim 5\times10^5$~s.
    }\label{fig:sf_dis_more_kepler2}
\end{figure*}

\begin{figure*}[t!]
    \centering
    \includegraphics[width=0.6\textwidth]{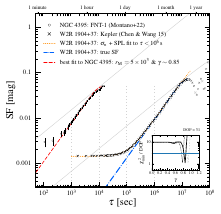}
    \includegraphics[width=0.4\textwidth]{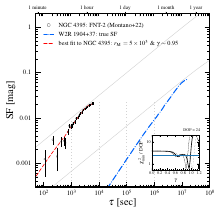}
    \includegraphics[width=0.4\textwidth]{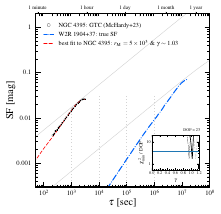}
        \caption{Same as Figure~\ref{fig:sf_dis}, but for comparing NGC 4395 ($M_{\rm BH} \simeq 10^4~M_\odot$) with W2R 1904+37 ($M_{\rm BH} \simeq 5 \times 10^7~M_\odot$) and $r_{\rm M} = 5 \times 10^3$. 
        The observed SF of W2R 1904+37 (crosses) is derived from the {\it Kepler} light curves stitched by \cite{chen2015}. The $\sigma_n$ + SPL fit is limited to $\uptau \lesssim 10^6$~s.
    }\label{fig:sf_dis_more_kepler3}
\end{figure*}

\begin{figure*}[t!]
    \centering
    
    \includegraphics[width=0.6\textwidth]{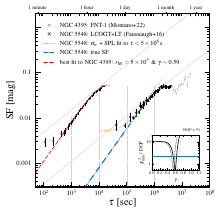}
    \includegraphics[width=0.4\textwidth]{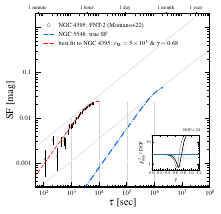}
    \includegraphics[width=0.4\textwidth]{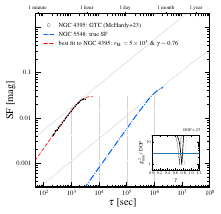}
    
    \caption{Same as Figure~\ref{fig:sf_dis}, but for comparing NGC 4395 ($M_{\rm BH} \simeq 10^4~M_\odot$) with NGC 5548 ($M_{\rm BH} \simeq 5 \times 10^7~M_\odot$) and $r_{\rm M} = 5 \times 10^3$. 
    The observed SF of NGC 5548 (crosses) is derived from the $i$-band light curve of \citet{Fausnaugh2016ApJ...821...56F}. The $\sigma_n$ + SPL fit is limited to $\uptau \lesssim 5 \times 10^5$~s.
    }\label{fig:sf_dis_more_5548}
\end{figure*}

Rather than considering a single typical damping timescale of the SF, directly comparing the global SFs of two AGNs to establish the dependence of timescales on BH mass are very informative, even with two AGNs only at a time. To do so, we should use their true SFs derived after having subtracted the noise contribution around the shortest timescales from the observed ones. 
For NGC 4395, its true SFs in three nights have been introduced in Section~\ref{sec:timerange_SF} and illustrated in the right panel of Figure~\ref{fig:3nights-sf} (see also circles in Figures~\ref{fig:sf_dis} to \ref{fig:sf_dis_more_5548}). 
For the four supermassive AGNs, especially Zw 229-15 and W2 1925+50 whose global SFs break at more than a damping timescale (Section~\ref{sec:sf_smbh}), we use a $\sigma_n$ + SPL function to fit their observed SFs at timescales smaller than $\uptau_2$, which broadly indicates the shortest breaking timescale of the SF (Table~\ref{tab:smbh}). Then, the corresponding true SFs are obtained after subtracting the best-fit $\sigma_n$ from the observed ones and are illustrated as the dot-dashed lines in Figures~\ref{fig:sf_dis} to \ref{fig:sf_dis_more_5548}.

To determine the timescale-mass scaling, $\uptau \propto r^\gamma_{\rm M}$, where $r_{\rm M}$ is the mass ratio of an AGN at the supermassive scale to NGC 4395 (cf. Tables~\ref{tab:smbh} and \ref{tab:mass_gamma}) and $\gamma$ is the mass index, we fit the true SF of an AGN at the supermassive scale to that of NGC 4395 at the intermediate scale. 
For a given $r_{\rm M}$, the true SF of an AGN at the supermassive scale is then scaled horizontally by a factor of $r_{\rm M}^\gamma$ for a series of $\gamma$ between 0 and 1.2, before fitting vertically to the true SF of NGC 4395. 
Note that for the GTC monitoring on NGC 4395, only data points at $\uptau\gtrsim 200$~s are used in the fit to avoid the clear residual noise (see the right panel of Figure~\ref{fig:3nights-sf}). 

For each $r_{\rm M}$ and $\gamma$, the fit results in a minimal reduced chi-square, $\chi^2_{\rm min}/{\rm DOF}$, where $\chi^2_{\rm min}$ is the minimal chi-square and DOF is the degree of freedom.
The insets of Figures~\ref{fig:sf_dis} to \ref{fig:sf_dis_more_5548} illustrate $\chi^2_{\rm min}/{\rm DOF}$ as a function of $\gamma$ for our fiducial $r_{\rm M}$ and two more mass ratios, i.e., $r_{\rm M}/2$ and $2 r_{\rm M}$. The fiducial $r_{\rm M}$ corresponds to the ratio of the measured BH masses (Table~\ref{tab:smbh}), while the other two mass ratios are considered for illustrating the effect induced by the measured uncertainty of the BH mass. For each $r_{\rm M}$, we determine a best $\gamma$ corresponding to the minimum of $\chi^2_{\rm min}(\gamma)/{\rm DOF}$, i.e., ${\rm Min}[\chi^2_{\rm min}/{\rm DOF}]$, and assess the associated 3$\sigma$ confidence level corresponding to ${\rm Min}[(\chi^2_{\rm min} + \Delta \chi^2)/{\rm DOF}]$, where $\Delta \chi^2$ is related to the DOF \citep[][cf. their Section~5.3.1]{Wall2012}. All fitting results are tabulated in Table~\ref{tab:mass_gamma}.

From Figures~\ref{fig:sf_dis} to \ref{fig:sf_dis_more_5548}, we compare the SFs of four AGNs at the supermassive scale to the three-night SFs of NGC 4395 at the intermediate-mass scale. Overall, the best $\gamma$ are all larger than 0.5 and up to $\sim 1.1$. 
On one hand, there is a degeneracy between $r_{\rm M}$ and $\gamma$, that is, the larger $r_{\rm M}$ the smaller $\gamma$. If the mass ratio increases by a factor of four, i.e., from $r_{\rm M}/2$ to $2~r_{\rm M}$, the resultant $\gamma$ decreases by $\sim 0.13$.
On the other hand, the best $\gamma$ surprisingly depends on the baseline of the light curve. 

For all four AGNs at the supermassive scale, the resultant $\gamma$ is larger by $\sim 0.16$ when the baseline of NGC 4395 decreases from 6.7 hr of FNT-1 to 2.2 hr of GTC. This strongly suggests that the larger $\gamma$ inferred from the comparison to a shorter baseline of NGC 4395 is likely a bias. A shorter baseline results in a smaller but unreal break timescale on the SF (cf. the $i$-band SFs of NGC 4395 in the right panel of Figure~\ref{fig:3nights-sf} and the corresponding best-fit $\tau_{\rm d}$ in Table~\ref{tab:fracs}).

\begin{figure}
    \centering
    \includegraphics[width=0.45\textwidth]{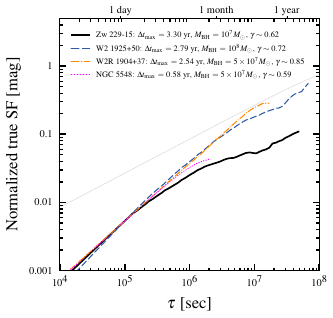}
    \caption{The true SFs of four AGNs at the supermassive scale, i.e., Zw 229-15 (solid curve), W2 1925+50 (dashed curve), W2R 1904+37 (dot-dashed curve), and NGC 5548 (dotted curve), are taken from Figures~\ref{fig:sf_dis} to \ref{fig:sf_dis_more_5548}, and normalized to that of Zw 229-15 at $\uptau = 10^5$~s. 
    For the three {\it Kepler} AGNs, the SF of Zw 229-15 flattens first, followed by W2 1925+50 and W2R 1904+37. 
    The BH mass of W2R 1904+37 intervenes between Zw 229-15 and W2 1925+50, however, its SF likely flattens at the longest timescale, indicating that its true BH mass may be larger than measured by \cite{Smith2018ApJ...857..141S}. 
    The baselines of Zw 229-15 and W2 1925+50 are the longest, and the resultant $\gamma$ are also comparable. 
    For NGC 5548, its baseline is too short, i.e., only 0.58 yr, to reach any firm conclusion on $\gamma$.
    }
    \label{fig:smbh_true_sf}
\end{figure}

Similarly, when comparing the SFs of AGNs at the supermassive scale to the FNT-1 SF of NGC 4395, a shorter baseline broadly gives rise to larger $\gamma$ (Table~\ref{tab:mass_gamma}). 
Figure~\ref{fig:smbh_true_sf} illustrates a direct comparison on the SFs of AGNs at the supermassive scale to highlight the effect of the limited baseline.
For Zw 229-15 and W2 1925+50, their baselines are almost the longest and the resultant $\gamma$ are comparable considering the fit uncertainties. 
Note that the reported BH mass of W2R 1904+37 is similar to that of NGC 5548 and smaller than that of W2 1925+50. However, the break timescale of the SF of W2R 1904+37 is clearly much larger than that of NGC 5548 and likely even larger than that of W2 1925+50. This suggests that the reported BH mass of W2R 1904+37 may have been underestimated. If the BH mass of W2 1925+50 were larger, the resultant $\gamma$ would be smaller and so likely be consistent with those inferred by using the SFs of Zw 229-15 and W2 1925+50. 
Last for NGC 5548, its baseline is too short to reach any firm conclusion, even the current fit gives a $\gamma$ comparable to those inferred by using the SFs of Zw 229-15 and W2 1925+50. 

Consequently, we suggest that the real $\gamma$ could be around $\sim 0.6 - 0.8$ according to the comparisons between the SFs of Zw 229-15 (and W2 1925+50) and the FNT-1/2 SF of NGC 4395.

Last but not least, we note that the observed TESS SF of NGC 4395 occupies a timescale range from $\sim 10^3$~s to $\sim 10^6$~s, even though its SF level is extremely lower than the FNT-1 SF (Figure~\ref{fig:sf_dis}). We make a tentative extension of the FNT-1 SF by extracting the ``true'' TESS SF and normalizing it to the true FNT-1 SF at $\sim 10^4$~s (open diamonds in Figure~\ref{fig:sf_dis}). Thus, the FNT-1 SF was extended to $\sim 10^6$~s. Astonishingly, we find a global consistency between the extended SF for NGC 4395 and the scaled SF for Zw 229-15 (with $\gamma \sim 0.62$). Note that the timescales at which the SF slope begins to departure from the DRW process, i.e., $\sim 0.3$ days for NGC 4395 and $\sim 10$ days for Zw 229-15, are consistent with that predicted by \cite{Sun2020ApJ...902....7S}, again suggesting a similar accretion process and variability in AGNs across different mass scales. 
In the future, directly extending the FNT-like observations on NGC 4395 from 1-2 days to $\sim$month would be valuable in learning the accretion physics.
Nevertheless, the SFs of AGNs at different mass scales could not be simply scalable, and any dependence or not of the slope on $M_{\rm BH}$ or other AGN parameters would be worthy of further investigation \citep{Simm2016A&A...585A.129S}. 

\subsubsection{A new timescale-mass scaling}

The timescale-mass scaling relation, $\uptau \propto M^\gamma$, has been extensively studied over the past decade. However, the mass indexes, $\gamma$, have not been agreed upon, and change from $\gamma < 0.5$ for a weak dependence, 
e.g., $\tau_{\rm d} \propto M^{\sim 0}_{\rm BH}$ \citep{Simm2016A&A...585A.129S}, 
$\tau_{\rm d} \propto M^{0.05 \pm 0.03}_{\rm BH} \lambda^{-0.09 \pm 0.02}_{\rm Edd}$ \citep{Suberlak2021ApJ...907...96S}, 
$\tau_{\rm d} \propto M^{0.14 \pm 0.12}_{\rm BH} \lambda^{-0.08 \pm 0.10}_{\rm Edd}$ \citep{MacLeod2010ApJ...721.1014M}, 
$\tau_{\rm d} \propto M^{0.38 \pm 0.15}_{\rm BH}$ \citep{Kozlowski2016ApJ...826..118K}, 
and
$\tau_{\rm d} \propto M^{0.38 \pm 0.05}_{\rm BH}$ \citep{,Burke2021Sci...373..789B},
to $\gamma > 0.5$ for a strong dependence, 
e.g., $\tau_{\rm d} \propto M^{0.61 \pm 0.47}_{\rm BH} \lambda^{-0.42 \pm 0.28}_{\rm Edd}$ \citep{Kelly2009ApJ...698..895K},
$\tau_{\rm d} \propto M^{0.62}_{\rm BH} \lambda^{0.48}_{\rm Edd}$ \citep{Guo2017ApJ...847..132G}, and
$\tau_{\rm d} \propto M^{0.55-0.65}_{\rm BH}\lambda^{0.35-0.3}_{\rm Edd}$ \citep{Arevalo2023}. 
Note all these works adopt the DRW damping timescale, $\tau_{\rm d}$.

Utilizing the {\it Zwicky Transient Facility} (ZTF) $g$-band light curves of 4770 quasars, \cite{Arevalo2023} grouped these light curves into 26 independent bins of $M_{\rm BH}$ and $\lambda_{\rm Edd}$, and calculated a low-resolution $g$-band power spectrum for each bin using Mexican hat filter \citep{arevalo2012}. 
They approximated these power spectra by using a straightforward analytical model, i.e., a bending power-law model, which incorporates a break timescale, $\tau_{\rm d}$, and two slopes below and above $\tau_{\rm d}$, and favored a steep slope below $\tau_{\rm d}$, indicating a departure from the DRW process. 
Besides, with $\tau_{\rm d} \propto M^{0.55-0.65}_{\rm BH}\lambda^{0.35-0.3}_{\rm Edd}$, \cite{Arevalo2023} found that all power spectra can be simplified to a universal power spectrum.
Their work depends on the assumed model with two slopes to account for the observed ZTF power spectra, which probe the variation at timescales of $\sim 3$ days to $300$ days. 
However, according to what we have discussed in Section~\ref{sec:sf_smbh}, the global SFs of AGNs may have two or more breaks, which are worthy of justification with the arrival new time-domain surveys.

Being different from all previous works, we do not use the single damping timescale ($\tau_{\rm d}$) but utilize the whole variation properties across nearly all available timescales.
Interestingly, our mass index agrees well with that derived by \cite{Arevalo2023} using a different method. 
Furthermore, our mass index is quite consistent with the prediction of the simple static standard thin disk, i.e., $\tau_{\rm d} \propto M^{0.5}_{\rm BH}$, assuming $\tau_{\rm d}$ is related to the thermal timescale, and is more consistent with $\tau_{\rm d} \propto M_{\rm BH}^{0.65 \pm 0.01} \lambda_{\rm Edd}^{0.65 \pm 0.01}$ \citep{zhou2024}, predicted by the corona heated accretion disc reprocessing model \citep{Sun2020ApJ...891..178S,Sun2020ApJ...902....7S}, which takes the time-dependent evolution of the standard thin disk into account. 

Additionally, it is interesting to note that \cite{Montano2022} reported a $\tau_{\rm gz}-M_{\rm BH}$ relation with a similar mass index of $0.63^{+0.16}_{-0.21}$, where $\tau_{\rm gz}$ is the time delay of the $z$-band variation relative to the $g$-band one. Although the lag $\tau_{\rm gz}$ is very different from the timescales of SFs we discussed here, the similar mass index may also suggest that they have the same physical origin.
Nonetheless, we caution that the $\tau_{\rm gz}-M_{\rm BH}$ relation could be subject to the same selection bias affecting the lag-luminosity relation, as discussed recently by \cite{Chen2024}.

\subsection{Color Variation (CV)}\label{sec:comp_cv}

\subsubsection{Plausible timescale-dependent CVs at the intermediate-mass scale}\label{sec:cv_all}

\begin{figure*}[t!]
    \centering    \includegraphics[width=0.9\textwidth]{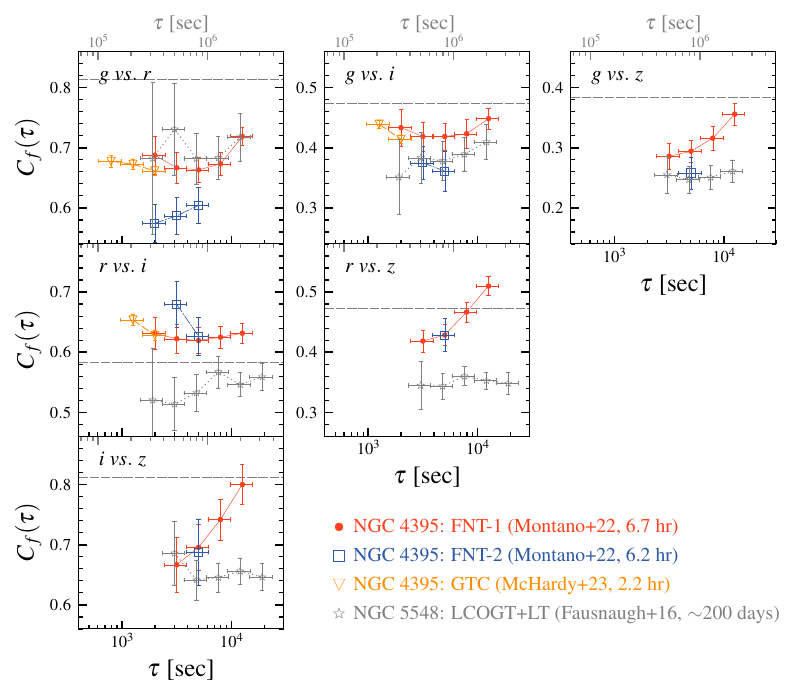}
    \caption{Comparison on the CVs in the flux space between $griz$ bands for AGNs across the intermediate-mass (NGC 4395 with $M_{\rm BH} \simeq 10^4~M_\odot$) to supermassive (NGC 5548 with $M_{\rm BH} \simeq 5 \times 10^7~M_\odot$) scales. 
    For NGC 4395, shown are CVs in their nights: the FNT-1 \citep[][filled circles]{Montano2022}, the FNT-2 \citep[][open squares]{Montano2022}, and the GTC night \citep[][open triangles]{McHardy2023MNRAS.519.3366M}.
    For NGC 5548, the $griz$-band light curves of \citet{Fausnaugh2016ApJ...821...56F} are used to derive the CVs (open asterisks with the top gray axes), timescales of which are scaled down by a factor of $r_{\rm M}^\gamma$ with $r_{\rm M} = 5000$ and $\gamma = 0.6$ (open asterisks with the bottom black axes).
    In each panel, the gray dashed horizontal line indicates the mean flux ratio of the redder band to the bluer one over three nights, ${\bar f^r}/{\bar f^b}$, below (above) which AGNs behave bluer (redder) when brightening. 
    Note that shown are only those data points within the proper range of timescales deemed to be less affected by both the photometric uncertainty and limited baseline (Figure~\ref{fig:i-z}).
    The $y$-axis of all panels has the same relative dynamical range.
    }
    \label{fig:cv_all}
\end{figure*}

Adopting the 3$\sigma$-cut variation significance and within the proper ranges of timescales for any two bands, Figure~\ref{fig:cv_all} shows in the flux space the CVs for NGC 4395 monitored in three nights: the FNT-1 (filled circles), the FNT-2 (open squares), and the GTC night (open triangles). 
The CVs vary much in the three nights of observation. 
The FNT-1 CVs based on the FNT-1 light curves span a widest range of timescale and the timescale dependency seems to be the most significant.
Although the FNT-2 CVs based on the FNT-2 light curves span a narrower range of timescale than the FNT-1 CVs, a nicely consistency between them can be found in several band pairs (e.g., $r-z$ and $i-z$), while a combination of them may suggest a clearer timescale dependence (e.g., $g-r$ and $g-i$).
Owing to the short baseline of the GTC light curves and the large GTC $z$-band photometric uncertainty (Table~\ref{tab:fracs}), there is no remaining data point for the GTC CVs involving the $z$ band and the GTC CVs are globally less reliable.

Since the FNT-1 CVs span the widest range of timescale and show a potential timescale dependency, we present a quantitative discussion on the timescale dependency of the FNT-1 CVs.
We employ a linear least-squares regression fitting package \texttt{scipy.stats.linregress} \citep{SciPy} to assess the significance of the timescale dependency with a smaller $p$-value indicating a more significant departure from a zero slope (i.e., no timescale dependence). 
As depicted in the diagonal sub-panels of Figure~\ref{fig:cv_all}, we confirm significant timescale-dependent CVs in the band pairs involving {\it z} band. The {\it p}-values are 0.04, 0.03, and 0.01 for the {\it g-z}, {\it r-z}, and {\it i-z} band pairs, respectively.
For the other band pairs (i.e., {\it g-r}, {\it g-i}, and {\it r-i}), we notice that these CVs decrease with increasing timescales up to $\sim 5000$~s (hereafter, a declining trend), and the prominence of the global timescale dependency is reduced, with {\it p}-values of 0.41, 0.47, and 0.90 for the {\it g-r}, {\it g-i}, and {\it r-i} band pairs, respectively. 

The CV approximately measures the ratio of SFs between two bands, and thus is more easily affected by poorer observational conditions and the randomness nature of AGN variability. 
First, the importance of the the baseline and photometric uncertainties of light curves is suggested by a coincidence: the most prominent CVs of NGC 4395 are revealed by the longest FNT-1 light curves.
Second, the effect of the randomness nature of AGN variability is demonstrated by another coincidence: clearer timescale-dependent CVs of NGC 4395 (Figure~\ref{fig:cv_all}) correspond to more significant variations with larger $\sigma_{\rm rms} / \sigma_{\rm e}$ (the right panel of Figure~\ref{fig:fvar}).

Besides, the contribution of the broad Balmer line can also reduce the timescale dependency. We notice that there are broad H$\alpha$ and H$\beta$ lines located in the $r$ and $g$ bands, respectively.
The variation of the broad H$\alpha$ line likely lags that of the $V$-band continuum by $\sim 50 - 100$~min \citep[][]{Woo2019NatAs...3..755W,Cho2020ApJ...892...93C,Cho2021ApJ...921...98C}, just corresponding to timescales of $\sim$ 3000~s to $\sim$ 6000~s. Being guided by simulations of \citet[][cf. their Figure~3]{Zhu2018}, besides the photometric uncertainty, the delayed H$\alpha$ (H$\beta$) variation in the $r$ ($g$) band can also contribute to the declining trend involved in the $r$ ($g$) versus $i$ CV. 
For the FNT-1 CVs of NGC 4395, if we exclude data points at timescales $\lesssim 5000$~s, very clear monotonic CVs are found for all band pairs. 

At any timescale, the CV always behaves BWB, except for the $r$ versus $i$ case (and somewhat the $r$ versus $z$ case), where RWB trends are found in all three nights. The RWB trend is just attributed to the fact that the $r$-band variation amplitudes are smaller than the $i$-band ones (the left panel of Figure~\ref{fig:fvar}), contrary to the general declining variation amplitude with increasing wavelength for AGNs. Although there is contribution of broad H$\alpha$ emission line in the $r$ band (Figure~\ref{fig:trans_spec}), the exact reason for the smaller $r$-band variation amplitude is unclear.

In short, using the current available data, we can only show plausible timescale-dependent CVs for NGC 4395 at the intermediate-mass scale.
Longer and repeated observations on NGC 4395 are essential for confirmation.

\subsubsection{Comparing CVs of AGNs across the intermediate-mass to supermassive scales}

Utilizing high-cadence ($\sim$ 1-3 days), long-baseline ($\sim 450$ days) {\it Swift} light curves in six UV/optical bands (i.e., UVW2, UVM2, UVW1, U, B, and V), \cite{Zhu2018} investigated the timescale dependency of CVs for NGC 5548. They revealed the timescale-dependent CVs in NGC 5548 at the supermassive scale for the first time. 
Taking the plausible timescale-dependent CVs for NGC 4395 at the intermediate-mass scale reported in Section~\ref{sec:cv_all}, we now have an unique opportunity to compare the CVs across different mass scales.
Before such a comparison, we first check the CVs for NGC 5548 in band pairs longer than those considered by \cite{Zhu2018}.

We notice that the {\it griz}-band light curves of NGC 5548 were obtained with ground-based telescopes only.
These light curves published by \cite{Fausnaugh2016ApJ...821...56F} span a baseline of $\sim$ 180 days much shorter than those of {\it Swift} used by \citet{Zhu2018}.
Following \citet{Zhu2018}, these light curves are binned with bin sizes of 0.5 days.
After that, we calculate the CVs in the flux space following the methodology outlined in Section~\ref{sec:cv}. The resultant CVs for NGC 5548 across the {\it griz} bands are depicted in Figure~\ref{fig:cv_all} as open asterisks with the top gray axes.
According to the linear regression analysis, the CVs of NGC 5548 involving {\it g-i} and {\it r-i} band pairs exhibit significant timescale dependency with positive slopes and {\it p}-values of 0.03 and 0.04, respectively. 
While for the {\it g-r}, {\it g-z}, and {\it r-z} band pairs, the significance are low with {\it p}-values of 0.81, 0.53, and 0.52, respectively. 
For the {\it i-z} band, a negative slope is even found with a {\it p}-value of 0.31. 

Shortening the light curves of NGC 5548 from $\sim 450$ days to $\sim 180$ days, we confirm that the CVs of the {\it Swift} band pairs are nearly the same as reported by \cite{Zhu2018}, except at the longest timescales around $\sim 100$ days where values of the CVs of almost all {\it Swift} band pairs become much smaller (i.e., behaving as a declining trend) for the shortened {\it Swift} light curves. 
This fact may suggest that for the {\it r-i}, {\it r-z}, and {\it i-z} band pairs the declining trends around $\sim 2 \times 10^6$~s found in Figure~\ref{fig:cv_all} are likely due to the limited baseline. 
Therefore, much longer and regular monitoring on NGC 5548 at optical wavelengths is necessary before claiming the timescale dependency of the CVs between optical bands.

Nevertheless, we make a tentative comparison on the CVs between NGC 5548 and NGC 4395. Assuming $\gamma = 0.6$ (Section~\ref{sec:comp_sf}), we horizontally scale the CVs of NGC 5548 by a factor of $r_{\rm M}^\gamma$ where $r_{\rm M} = 5000$.
The scaled CVs of NGC 5548 (represented by open asterisks with the bottom black axes) are depicted in Figure~\ref{fig:cv_all}, revealing a similar dynamical range of timescales showing the timescale dependency for both NGC 5548 and NGC 4395.
In the future, with longer light curves we can better measure the CVs for AGNs across different mass scales and could use them to directly compare and constrain the mass index $\gamma$.

\subsection{New evidence against the traditional reprocessing scenario but for the thermal fluctuation scenario?}

\begin{figure}[t!]
\includegraphics[width=0.45\textwidth]{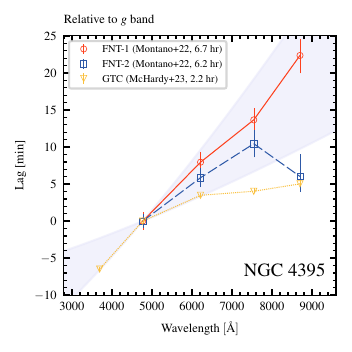}
\caption{Three distinct lag-wavelength relations for NGC 4395 measured in three nights by \citet[][2.2 hr]{McHardy2023MNRAS.519.3366M} and \citet[][6.7 hr for the FNT-1 and 6.2 hr for the FNT-2]{Montano2022} plausibly indicate the randomness of lag implied by the disk thermal fluctuation \citep{Cai2018,Cai2020}. The time lags are relative to the $g$ band. The shaded region is enclosed by two SSD-predicted lag-wavelength relations calculated following Equation (12) of \cite{Fausnaugh2016ApJ...821...56F} with the radiation efficiency $\eta=0.1$, the ratio of external to internal heating $\kappa = 1$, the conversion factor of temperature to wavelength $X=5.04$ \citep{tie2018}, the bolometric luminosity of $5.3\times10^{40}~{\rm erg~s^{-1}}$ \citep{Moran2005}, and two distinct BH masses of $1.7\times10^4~M_\odot$ (\citealt{Cho2021ApJ...921...98C}) and $4\times10^5~M_\odot$ (\citealt{denBrok2015ApJ...809..101D}).}
\label{fig:lag}
\end{figure}

By analyzing multi-band light curves of NGC 4395, all studies \citep{Desroches2006,McHardy2016,Montano2022, McHardy2023MNRAS.519.3366M} had merely focused on its lag-wavelength relation, claiming either a consistency with or a departure from the prediction by the \citet{Shakura&Sunyaev1973} disk (SSD; Figure~\ref{fig:lag}). 
The diverse lag-wavelength relations measured in different nights for NGC 4395 had been interpreted differently: randomness of thermal fluctuation (\citealt{Cai2020}; Z. B. Su et al. 2024a, in preparation), contamination of the diffuse continuum emission from the broad-line region \citep{Montano2022}, or the imprint of the outer edge of the accretion disk \citep{McHardy2023MNRAS.519.3366M}.

Here, we propose the diverse lag-wavelength relations for NGC 4395 (Figure~\ref{fig:lag}) as a potential new evidence against the traditional reprocessing scenario, which implies very stable lags \citep[][cf. their Figure~3]{Sun2019MNRAS} and cannot explain the large and significant difference in lags among nights.
In Figure~\ref{fig:lag}, for the $r$, $i$, and $z$ bands relative to the $g$ band, the ratios of lags measured by \citet[][the Night-1]{Montano2022} and those by \citet{McHardy2023MNRAS.519.3366M} are as large as $\simeq 2.2$, $\simeq 3.3$, and $\simeq 4.2$ with confidence levels reaching $\simeq 3.2\sigma$, $\simeq 5.8\sigma$, and $\simeq 7.0\sigma$, respectively.
Admittedly, the mean fluxes of \citet{McHardy2023MNRAS.519.3366M} are lower than those of \citet{Montano2022} by a factor of $\simeq 1.14$ on average (Figure~\ref{fig:lc}). According to the lag-luminosity relation, i.e., $\tau_{\rm lag}\propto L^{\rm 0.5}$ \citep{sergeev2005}, the smaller flux by a factor of $\simeq 1.14$ corresponds to a smaller lag by a factor of $\simeq 1.07$, which is too small to account for the large difference in the observed lags.

Instead, a large difference in lags as a result of the randomness of disk fluctuation has been predicted by the simple thermal fluctuation scenario (\citealt{Cai2018,Cai2020}; Z. B. Su et al., 2024a, in preparation) and thermal fluctuations induced by corona heating through magnetic field \citep{Sun2020ApJ...891..178S,Sun2020ApJ...902....7S,Sun2023MNRAS.521.2954S}. More quantitative comparisons between the models (both the thermal fluctuation model and the reprocessing model) and the data (both the lag-wavelength relation and the timescale-dependent CV) are being conducted for NGC 4395 (Z. B. Su et al. 2024c, in preparation) after we have upgraded the radiation efficiency for general AGNs, including NGC 4395 (briefly, its $\eta$ is likely as small as $\sim 0.001$ rather than the commonly assumed 0.1; Z. B. Su et al. 2024b, in preparation).

\section{Summary}\label{sec:conclusion}

In this work, we analyze three nights of {\it griz}-band light curves of NGC 4395. At the intermediate-mass scale, we unveil for the first time prominent departures from the DRW process at timescales shorter than $\sim$ hours in all three nights and four bands of observation on NGC 4395. Additionally, we can only show plausible timescale-dependent CVs for NGC 4395 in the longest two nights of observation. 

By comparing optical variations, i.e., the SFs, across the intermediate-mass (NGC 4395) to supermassive (mainly Zw 229-15 and W2 1925+50) scales, we suggest a new scaling relation between the optical variation timescale and BH mass with a mass slope of $\simeq 0.6-0.8$. This mass index differs from most previous measurements but is consistent with a recent theoretical prediction by the time-dependent evolution of the standard thin disk, suggesting a similar accretion process and variability in AGNs across various mass scales.

In near future, many facilities, such as the Multi-channel Imager onboard the Chinese Space Station Telescope (\citealt{csst}), the Multi-channel Photometric Survey telescope \citep[Mephisto; e.g.,][]{Liu+etal+2019, Lei+etal+2020}, and the Wide Field Survey Telescope \citep{WFST2023SCPMA..6609512W} will be extremely helpful in exploring the AGN variability, including the SF, CV, and lag-wavelength relation.

\section*{Acknowledgement}

We are grateful to the anonymous referee for many constructive comments, to Ian McHardy for sharing and clarifying the light curves of NGC 4395, and to Xue-Bing Wu for helpful comments.
This work is supported by National Key R\&D Program of China (grant No. 2023YFA1607903) and the National Science Foundation of China (grant Nos. 12373016 and 12033006).
Z.Y.C. acknowledges support from the science research grants from the China Manned Space Project under grant no. CMS-CSST-2021-A06 and the Cyrus Chun Ying Tang Foundations.
M.Y.S. acknowledges support from the National Natural Science Foundation of China (NSFC-12322303) and the Natural Science Foundation of Fujian Province of China (No. 2022J06002).

\software{\julia\ Programming Language \citep{Julia2017}, \texttt{UncertainData.jl} \citep{Haaga2019}, \texttt{Numpy} \citep{Numpy}, \texttt{Scipy} \citep{SciPy}, \texttt{Matplotlib} \citep{Matplotlib}.}

\bibliographystyle{aasjournal}
\bibliography{ms.bbl}

\end{document}